\documentclass[12pt]{article}
\pdfoutput=1

\usepackage[a4paper,text={16.8cm,22.4cm}]{geometry}
\usepackage{amsmath,amsfonts,slashed,amssymb,tikz,bm,psfrag,graphicx,color,dsfont}
\usepackage{hyperref}
\hypersetup{
    colorlinks=true,
    linkcolor=black,
    citecolor=black,
    filecolor=black,
    urlcolor=black,
}
\RequirePackage[sort&compress,square,comma,numbers]{natbib}
\allowdisplaybreaks 
\renewcommand{\arraystretch}{1.2}

\begin{document}

\begin{titlepage}

\begin{flushright}
\normalsize
MITP/16-070\\
September 20, 2016
\end{flushright}

\vspace{0.1cm}
\begin{center}
\Large\bf
Exclusive Weak Radiative Higgs Decays in the Standard Model and Beyond
\end{center}

\vspace{0.5cm}
\begin{center}
Stefan Alte$^a$, Matthias K\"onig$^a$ and Matthias Neubert$^{a,b}$\\
\vspace{0.7cm} 
{\sl ${}^a$PRISMA Cluster of Excellence \& Mainz Institute for Theoretical Physics\\
Johannes Gutenberg University, 55099 Mainz, Germany\\[3mm]
${}^b$Department of Physics, LEPP, Cornell University, Ithaca, NY 14853, U.S.A.}
\end{center}

\vspace{0.8cm}
\begin{abstract}
We perform a detailed study of the exclusive Higgs decays $h\to MZ$ and $h\to MW$, where $M$ is a pseudoscalar or vector meson, using the QCD factorization approach. We allow for the presence of new-physics effects in the form of modified Higgs couplings to gauge bosons and fermions, including the possibility of flavor-changing Higgs couplings. We show that the decays $h\to VZ$ exhibit a strong sensitivity to the effective CP-even and CP-odd $h\gamma Z$ couplings. When combined with a measurement of the $h\to\gamma Z$ decay rate, this can be used to extract these couplings up to a sign ambiguity in the CP-odd coefficient. Some of the $h\to MW$ decay modes can be used to probe for flavor-violating Higgs couplings involving the top quark.
\end{abstract}
\vfil

\end{titlepage}

\section{Introduction}
\label{sec:intro}

After the discovery of the Higgs boson \cite{Aad:2012tfa,Chatrchyan:2012ufa}, many questions regarding its properties remain unanswered. In the Standard Model (SM), the Higgs boson couples to fermions through Yukawa interactions, which after electroweak symmetry breaking generate the fermion masses. These masses, which exhibit a large hierarchy spanning many orders of magnitude, enter as input parameters in the SM Lagrangian, and the question whether their hierarchy follows a more fundamental pattern remains open. The fermion mass matrix is aligned with the Yukawa couplings, and as a result the Higgs couplings are flavor-diagonal in the physical basis. Models beyond the SM exist, in which the Higgs couplings to fermions are changed in a non-trivial way, potentially leading to sizeable deviations from the SM predictions and allowing for flavor-changing and CP-violating Higgs interactions. Additionally, heavy new particles can induce non-standard Higgs couplings to gauge bosons. At present, the loop-induced $h\gamma Z$ couplings are least constrained by experimental data. It is of paramount importance to probe these couplings in any way conceivable.

In this work we advocate the use of the exclusive weak radiative Higgs-boson decays $h\to M V$, where $M$ denotes a meson and $V=Z,W$ an electroweak gauge boson, as probes for non-standard Higgs couplings. The case $V=\gamma$ has already been studied in great detail in \cite{Isidori:2013cla,Bodwin:2013gca,Kagan:2014ila,Bodwin:2014bpa,Koenig:2015pha}. The corresponding decay amplitudes receive contributions from two types of decay topologies, which interfere destructively: The ``direct contributions'', which involve the coupling of the Higgs boson to the quarks forming the meson, and the ``indirect contributions'', in which the Higgs decays to an off-shell vector boson that converts to the meson through a local matrix element. The direct amplitudes can be evaluated in the framework of the QCD factorization approach \cite{Lepage:1979zb,Lepage:1980fj,Efremov:1978rn,Efremov:1979qk,Chernyak:1983ej}, in which the large separation between the hard scattering scale $m_h$ and the hadronic scale $\Lambda_\mathrm{QCD}$ yields to a factorization of the amplitudes into convolutions of hard functions with light-cone distribution amplitudes (LCDAs) for the meson $M$. While the hard function can be calculated in perturbation theory, the LCDAs encode the physics at the hadronic scale and have to be extracted from non-perturbative methods such as lattice gauge theory or QCD sum rules. For mesons containing heavy quarks, insight into the structure of the LCDAs can be obtained using heavy-quark effective theories such as NRQCD and HQET. The QCD factorization formula can be derived elegantly using soft-collinear effective theory \cite{Bauer:2000yr,Bauer:2001yt,Bauer:2002nz,Beneke:2002ph}, as has been demonstrated in \cite{Grossmann:2015lea,Alte:2015dpo}.

The interplay between direct and indirect contributions gives rise to a strong sensitivity of the $h\to M\gamma$ decay rates on the quark Yukawa couplings. While it is challenging to reconstruct these rare decays at the LHC \cite{Perez:2015lra}, it should be possible to significantly improve existing searches \cite{Aad:2015sda,Khachatryan:2015lga,Aaboud:2016rug} in the high-luminosity phase at the LHC. The aim of the present work is to investigate whether such an interference pattern persists in the case of the weak radiative decays $h\to M Z$ and $h\to M W$, and what other possibilities of probing new-physics effects open up in these modes. Some of these modes have already been explored in the literature. The authors of \cite{Isidori:2013cla,Gonzalez-Alonso:2014rla} have discussed the indirect contributions to the $h\to MZ$ and $h\to MW$ decay amplitudes induced by the $hZZ$ and $hWW$ vertices. They have missed the indirect contributions involving the effective $h\gamma Z$ vertex, which turn out to give the dominant effects for $h\to VZ$ decay modes containing a light final-state vector meson. We also extend their work in an important way by performing a careful treatment of the flavor-specific decay constants of neutral mesons, and by evaluating the direct contributions to the amplitudes and studying to which extent these are suppressed. In \cite{Gao:2014xlv} and \cite{Modak:2014ywa} the decays $h\to J/\psi\,Z$ and $h\to\Upsilon(1S)\,Z$ into heavy quarkonia have been analyzed including the indirect contributions involving both the $hZZ$ and $h\gamma Z$ vertices. The second paper also provides an estimate of the direct contributions based on the non-relativistic approximation. The sign of the interference term of the two indirect contributions found by these authors appears to be opposite to the one we obtain, and as a result their branching ratios are typically about 40\% larger than our values. The authors of \cite{Kagan:2014ila} have very briefly discussed the sensitivity of the $h\to B^* W$ decay rate to the flavor-changing Higgs couplings to top and up quarks. Our detailed analysis does not fully confirm the result presented in this paper.

In our analysis we assume SM couplings for all particles other than the Higgs boson. For the Higgs interactions with SM particles, we use the phenomenological Lagrangian
\begin{equation}\label{eqn:effL}
\begin{aligned}
   {\cal L}_{\rm eff} &= \kappa_W\,\frac{2m_W^2}{v}\,h\,W_\mu^+ W^{-\mu} 
    + \kappa_Z\,\frac{m_Z^2}{v}\,h\,Z_\mu Z^\mu 
    - \frac{h}{\sqrt2}\,\sum_{f=u,d,e} \Big( \bar f_L Y_f f_R + \mbox{h.c.} \Big) \\[-2mm]
   &\quad\mbox{}+ \frac{\alpha}{4\pi v} \left( 
    \kappa_{\gamma\gamma}\,h\,F_{\mu\nu} F^{\mu\nu}
    - \tilde\kappa_{\gamma\gamma}\,h\,F_{\mu\nu} \tilde F^{\mu\nu} 
    + \frac{2\kappa_{\gamma Z}}{s_W c_W}\,h\,F_{\mu\nu} Z^{\mu\nu} 
    - \frac{2\tilde\kappa_{\gamma Z}}{s_W c_W}\,h\,F_{\mu\nu} \tilde Z^{\mu\nu} \right) ,
\end{aligned}
\end{equation}
where $s_W\equiv\sin\theta_W$ and $c_W\equiv\cos\theta_W$ are the sine and cosine of the weak mixing angle. Here $Y_f$ are complex $3\times 3$ matrices in generation space. We normalize the flavor-diagonal entries of these matrices to the SM Yukawa couplings and define corresponding rescaling parameters $\kappa_{f_i}$ and $\tilde\kappa_{f_i}$ via
\begin{equation}
   \left(Y_f\right)_{ii} = \left( \kappa_{f_i} + i\tilde\kappa_{f_i} \right) \frac{\sqrt2\,m_{f_i}}{v} \,.
\end{equation}
For notational convenience, we will use the name of a given fermion instead of the label $f_i$ whenever possible. For the flavor off-diagonal Higgs couplings to quarks with $q_i,q_j\ne t$, the global analysis of indirect constraints performed in \cite{Harnik:2012pb} gives $|\!\left(Y_q\right)_{ij}\!|<10^{-5}\!-\!10^{-3}$. These couplings are so small that they will play no role in our analysis. The flavor-changing couplings to top quarks are constrained by LHC measurements of the branching ratios ${\rm Br}(t\to qh)$, where $q=c,u$. The most up-to-date bounds have been determined in \cite{Buschmann:2016uzg}, yielding (at 95\% confidence level)
\begin{equation}\label{eq:kappaconstr}
   \sqrt{\left|Y_{tc}\right|^2+\left|Y_{ct}\right|^2} < 0.18 \,, \qquad 
   \sqrt{\left|Y_{tu}\right|^2+\left|Y_{ut}\right|^2} < 0.17
\end{equation}
at the scale $\mu=m_h$. For an integrated luminosity of 3000\,fb$^{-1}$ at $\sqrt{s}=14$\,TeV, the bounds are expected to improve to 0.04 in both cases \cite{Gorbahn:2014sha}.

\section{\boldmath Weak radiative hadronic decays $h\to MZ$}
\label{sec:hMZ}

\begin{figure}
\begin{center}
\raisebox{0pt}{\includegraphics[width=0.24\textwidth]{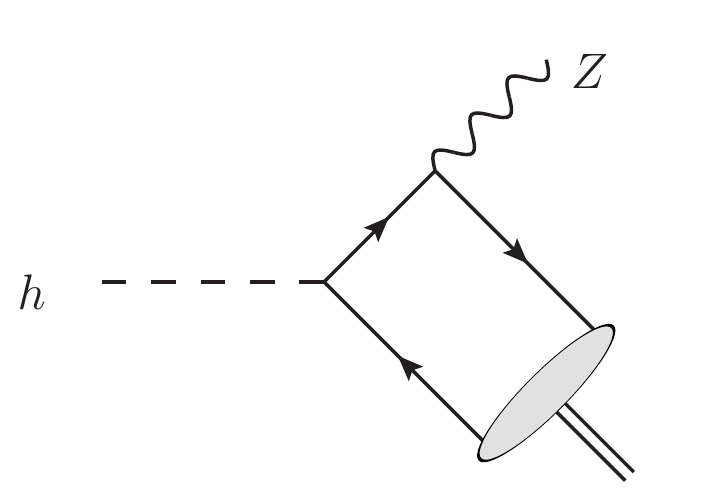}}
\raisebox{-5pt}{\includegraphics[width=0.24\textwidth]{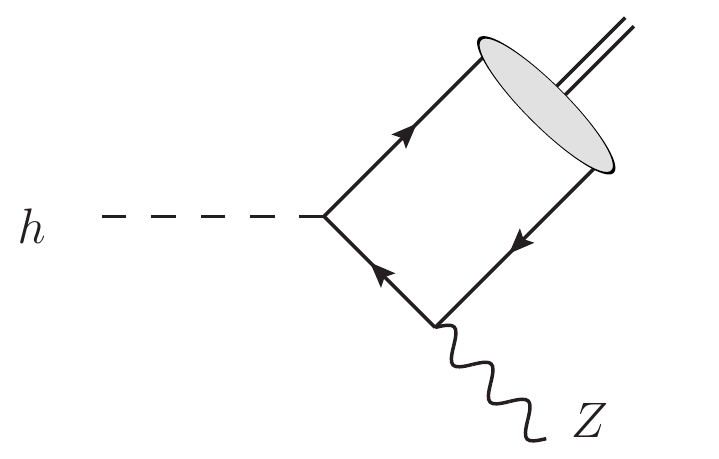}}
\raisebox{0pt}{\includegraphics[width=0.24\textwidth]{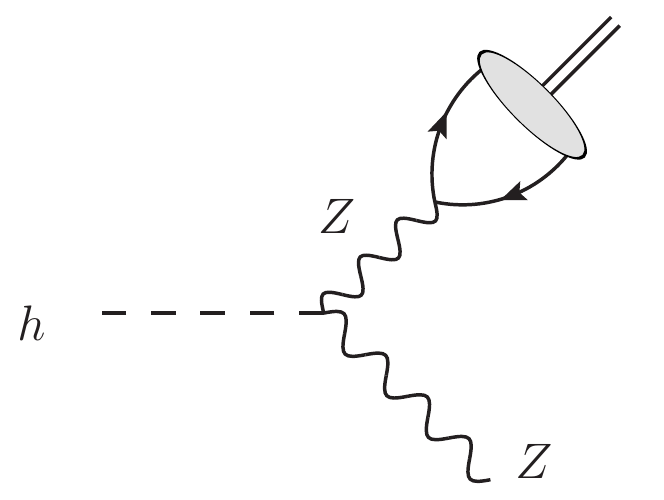}}
\raisebox{0pt}{\includegraphics[width=0.24\textwidth]{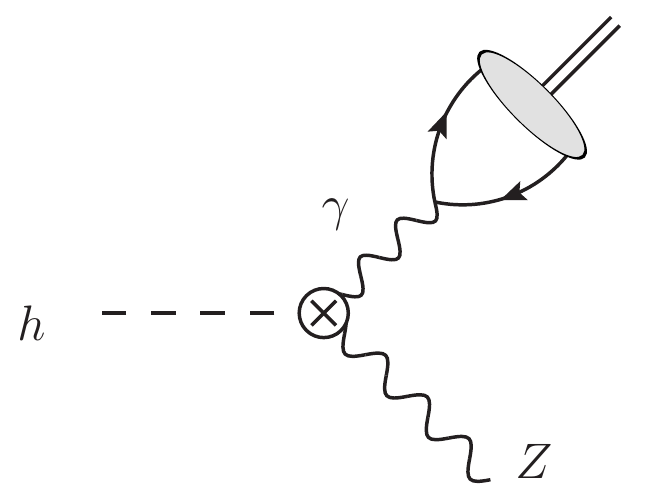}}
\parbox{15.5cm}
{\caption{\label{fig:hMZ}
Leading-order diagrams contributing to the decays $h\to M Z$. The crossed circle in the last graph denotes contributions from one-loop SM diagrams to $h\to Z\gamma^*$ as well as tree-level effective vertices. This last graph only contributes if $M=V_\perp$.}}
\end{center}
\end{figure}

The decays $h\to MZ$ are interesting by the fact that the massive final-state gauge boson can be in a longitudinal polarization state. As a consequence, both pseudoscalar and vector mesons can be produced, whereas in the case of $h\to M \gamma$ decays $M$ could only be a (transversely polarized) vector meson \cite{Kagan:2014ila,Bodwin:2014bpa,Koenig:2015pha}. The relevant Feynman diagrams for the decays $h\to MZ$ are depicted in Figure~\ref{fig:hMZ}. The first two graphs show the direct contributions to the decay amplitude at the leading order. In these diagrams, the Higgs boson couples to the quark and anti-quark pair from which the meson is formed. The indirect contributions to the decay amplitude are shown by the last two diagrams, in which the Higgs boson decays into a $ZZ^*$ or $Z\gamma^*$ boson pair followed by the decay of the off-shell boson into the final-state meson. While the $hZZ$ vertex exists at tree level, the $h\gamma Z$ vertex is induced at one-loop order in the SM. Possible new-physics contributions to this vertex are parameterized by the operators $h\,F_{\mu\nu} Z^{\mu\nu}$ and $h\,F_{\mu\nu}\tilde Z^{\mu\nu}$ in the effective Lagrangian \eqref{eqn:effL}. We include both of these contributions in the effective vertex denoted by the crossed circle.

\subsection{Form factor decomposition}

The most general parameterization of the decay amplitudes into pseudoscalar mesons can be written as
\begin{equation}\label{hPZampl}
   i {\cal A}(h\to PZ)
   = \frac{2g}{c_W v}\,k\cdot\varepsilon_Z^*\,F^{PZ} \,,
\end{equation}
where $k$ denotes the meson momentum. The $Z$ boson is longitudinally polarized. The most general parameterization of the decay amplitudes into vector mesons can be chosen as
\begin{equation}\label{hVperpZampl}
   i{\cal A}(h\to VZ) 
   = - \frac{2g m_V}{c_W v} 
    \left[ \varepsilon_V^{\parallel *}\cdot\varepsilon_Z^{\parallel *}\,F_\parallel^{VZ} 
    + \varepsilon_V^{\perp *}\cdot\varepsilon_Z^{\perp *}\,F_\perp^{VZ} 
    + \frac{\epsilon_{\mu\nu\alpha\beta}\,k^\mu q^\nu\varepsilon_V^{*\alpha}\varepsilon_Z^{*\beta}}%
           {\left[ (k\cdot q)^2-k^2 q^2\right]^{1/2}}\,\widetilde F_\perp^{VZ} \right] ,
\end{equation}
where $q$ is the momentum of the $Z$ boson, 
\begin{equation}\label{hVpolarizations}
   \varepsilon_V^{\parallel \mu} = \frac{1}{m_V}\,\frac{k\cdot q}{\left[ (k\cdot q)^2-k^2 q^2\right]^{1/2}}
    \left( k^\mu - \frac{k^2}{k\cdot q}\,q^\mu \right) , \qquad
   \varepsilon_V^{\perp\mu} = \varepsilon_V^\mu - \varepsilon_V^{\parallel \mu}
\end{equation}
are the longitudinal and transverse polarization vectors of the vector meson, and analogous expressions (with $m_V\to m_Z$ and $k\leftrightarrow q$) apply for the polarization vectors of the $Z$ boson. The decay rates are found to be
\begin{equation*}
   \Gamma(h\to PZ) 
   = \frac{m_h^3}{4\pi v^4}\,\lambda^{3/2}(1,r_Z,r_P)\,\big| F^{PZ} \big|^2 \,, \hspace{4.1cm}
\end{equation*}
\begin{equation}\label{hMZrates}
\begin{aligned}
   \Gamma(h\to VZ) 
   &= \frac{m_h^3}{4\pi v^4}\,\lambda^{1/2}(1,r_Z,r_V)\,(1-r_Z-r_V)^2 \\
   &\quad\times \left[
    \big| F_\parallel^{VZ} \big|^2 + \frac{8r_V r_Z}{(1-r_Z-r_V)^2}
    \Big( \big| F_\perp^{VZ} \big|^2 + \big| \widetilde F_\perp^{VZ} \big|^2 \Big) \right] ,
\end{aligned}
\end{equation}
where $\lambda(x,y,z)=(x-y-z)^2-4yz$, and we have defined the mass ratios $r_Z=m_Z^2/m_h^2$ and $r_M=m_M^2/m_h^2$. Notice that the decay rates into transversely polarized vector mesons are suppressed, relative to the other rates, by a factor $r_V$. The mass ratios $r_P$ and $r_V$ are very small for all mesons considered in this work; however, it turns out that the contribution of the transverse polarization states to the $h\to VZ$ rates are significant, especially for light vector mesons. We will thus keep the dependence on all masses in our analysis.

The form factors in \eqref{hMZrates} contain the direct and indirect contributions. We start with the indirect contributions, since they are found to give rise to the dominant effects. They involve hadronic matrix elements of local currents and hence can be calculated to all orders in QCD. We obtain
\begin{equation}\label{AVindir}
\begin{aligned}
   F_{\rm indirect}^{PZ} &= \kappa_Z \sum_q f_P^q\,a_q \,, \\
   F_{\parallel\,\rm indirect}^{VZ} &= \frac{\kappa_Z}{1-r_V/r_Z} \sum_q f_V^q\,v_q 
    + C_{\gamma Z}\,\frac{\alpha(m_V)}{4\pi}\,\frac{4r_Z}{1-r_Z-r_V} \sum_q f_V^q\,Q_q \,, \\
   F_{\perp\,\rm indirect}^{VZ} &= \frac{\kappa_Z}{1-r_V/r_Z} \sum_q f_V^q\,v_q 
    + C_{\gamma Z}\,\frac{\alpha(m_V)}{4\pi}\,\frac{1-r_Z-r_V}{r_V} \sum_q f_V^q\,Q_q \,, \\
   \widetilde F_{\perp\,\rm indirect}^{VZ} 
   &= \widetilde C_{\gamma Z}\,\frac{\alpha(m_V)}{4\pi}\,
    \frac{\lambda^{1/2}(1,r_Z,r_V)}{r_V} \sum_q f_V^q\,Q_q \,,
\end{aligned}
\end{equation}
where $v_q=T_3^q/2-Q_q s_W^2$ and $a_q=T_3^q/2$ are the vector and axial-vector couplings of the $Z$ boson to the quark $q$. The flavor-specific decay constants $f_P^q$ and $f_V^q$ are defined in terms of the local matrix elements 
\begin{equation}\label{fPqdef}
   \langle P(k)|\,\bar q\,\gamma^\mu\gamma_5\,q\,|0\rangle = -if_P^q\,k^\mu \,, \qquad
   \langle V(k,\varepsilon)|\,\bar q\,\gamma^\mu q\,|0\rangle = -if_V^q\,m_V\,\varepsilon^{*\mu} \,,
\end{equation}
with $q=u,d,s,\dots$. These quantities arise because flavor-diagonal neutral mesons must in general be described as superpositions of valence quark-antiquark states with different flavor. The coefficients $C_{\gamma Z}$ and $\widetilde C_{\gamma Z}$ are given by (with $\tau_i=4m_i^2/m_h^2$) \cite{Bergstrom:1985hp}
\begin{align}\label{Cgaga}
   C_{\gamma Z} 
   &= \sum_q \kappa_q\,\frac{2 N_c Q_q v_q}{3}\,A_f(\tau_q,r_Z)
    + \sum_l \kappa_l\,\frac{2 Q_l v_l}{3}\,A_f(\tau_l,r_Z) 
    - \frac{\kappa_W}{2}\,A_W^{\gamma Z}(\tau_W,r_Z) + \kappa_{\gamma Z} \,, \nonumber\\
   \widetilde C_{\gamma Z} 
   &= \sum_q \tilde\kappa_q\,N_c Q_q v_q\,B_f(\tau_q,r_Z)
    + \sum_l \tilde\kappa_l\,Q_l v_l\,B_f(\tau_l,r_Z)
    + \tilde \kappa_{\gamma Z} \,.
\end{align}
The loop functions $A_f$, $A_W^{\gamma Z}$ and $B_f$ are given in Appendix~D of \cite{Koenig:2015pha}. Note that $A_f$ and $B_f$ are strongly suppressed for all fermions except the top quark. QCD corrections to the $h\to Z\gamma$ amplitude were calculated in \cite{Spira:1991tj} and found to be very small, about 0.1\%. To evaluate these expressions we use the running quark masses evaluated at the hard scale $\mu_{hZ}=(m_h^2-m_Z^2)/m_h\approx 58.6$\,GeV, corresponding to twice the energy of the meson $M$ in the rest frame of the decaying Higgs boson, in the limit where the meson mass is neglected. We use the running quark masses at next-to-next-to-leading order (NNLO) in the $\overline{\rm MS}$ scheme, starting from the low-energy values given in \cite{Agashe:2014kda}. This yields $m_b(\mu_{hZ})=2.98$\,GeV, $m_c(\mu_{hZ})=664$\,MeV, $m_s(\mu_{hZ})=56.4$\,MeV, $m_d(\mu_{hZ})=2.84$\,MeV and $m_u(\mu_{hZ})=1.30$\,MeV. For the top quark we use the pole mass $m_t=173.34$\,GeV. Numerically, we obtain
\begin{equation}
\begin{aligned}
   C_{\gamma Z} &= \kappa_{\gamma Z} - 2.53\,\kappa_W + 0.135\,\kappa_t
    - (1.66 - 0.83\,i)\cdot 10^{-3}\,\kappa_b - (1.35 - 0.46\,i)\cdot 10^{-4}\,\kappa_c \\
   &\quad\mbox{}- (7.45 - 3.21\,i)\cdot 10^{-5}\,\kappa_\tau 
    - (1.35 - 0.30\,i)\cdot 10^{-6}\,\kappa_s + \dots 
    \,\stackrel{\rm SM}{\longrightarrow}\, -2.395 + 0.001\,i \,, \\
   \widetilde C_{\gamma Z} &= \tilde\kappa_{\gamma Z} + 0.206\,\tilde\kappa_t
    - (1.90 - 0.83\,i)\cdot 10^{-3}\,\tilde\kappa_b - (1.48 - 0.46\,i)\cdot 10^{-4}\,\tilde\kappa_c \\
   &\quad\mbox{}- (8.36 - 3.21\,i)\cdot 10^{-5}\,\tilde\kappa_\tau 
    - (1.43 - 0.30\,i)\cdot 10^{-6}\,\tilde\kappa_s + \dots 
    \,\stackrel{\rm SM}{\longrightarrow}\, 0 \,.
\end{aligned}
\end{equation}
Note that the contributions from light quarks and leptons in the loop are strongly suppressed, even if we allow for strongly enhanced Yukawa couplings of these fermions. Since the decay $h\to b\bar b$ is the dominant Higgs decay mode in the SM, the present phenomenological information about Higgs decays from the LHC suggests that $|\kappa_b|=O(1)$, while the Yukawa couplings of lighter fermions should not be larger than the $b$-quark Yukawa (see e.g.\ \cite{Kagan:2014ila,Perez:2015aoa}). This implies $|\kappa_\tau|\lesssim O(2)$, $|\kappa_c|\lesssim O(4)$, $|\kappa_s|\lesssim O(50)$. Similar bounds apply to the CP-odd coefficients $\tilde\kappa_f$. Even if these bounds were saturated this would have a very minor impact in the values of $C_{\gamma Z}$ and $\widetilde C_{\gamma Z}$. In our phenomenological analysis we will use the approximations
\begin{equation}\label{keffdef}
   C_{\gamma Z} = -2.395 + \kappa_{\gamma Z}^{\rm eff} \,, \qquad
   \widetilde C_{\gamma Z} = \tilde\kappa_{\gamma Z}^{\rm eff} \,,
\end{equation}
where the tiny imaginary parts can be safely neglected. The coefficients $\kappa_{\gamma Z}^{\rm eff}$ and $\tilde\kappa_{\gamma Z}^{\rm eff}$ parameterize new-physics effects and vanish in the SM. To a good approximation $\kappa_{\gamma Z}^{\rm eff}\approx\kappa_{\gamma Z}-2.53\,(\kappa_W-1)+0.135\,(\kappa_t-1)$ and $\tilde\kappa_{\gamma Z}^{\rm eff}\approx\tilde\kappa_{\gamma Z}+0.206\,\tilde\kappa_t$. The values of these two coefficients are currently not much constrained by data, because the decay $h\to\gamma Z$ has not yet been observed at the LHC. The current limits from CMS \cite{Chatrchyan:2013vaa} and ATLAS \cite{Aad:2014fia} imply upper bounds on the decay rates of 9 and 11 times the SM value, respectively, both at 95\% confidence level. The stronger bound from CMS implies the constraint
\begin{equation}\label{gaZrange}
   \sqrt{ \left| \kappa_{\gamma Z}^{\rm eff} - 2.395 \right|^2
    + \left| \tilde\kappa_{\gamma Z}^{\rm eff}\right|^2 }
   < 7.2 \,.
\end{equation}
A non-vanishing $\tilde\kappa_{\gamma Z}^{\rm eff}$ can induce contributions to the electric dipole moments (EDMs) of leptons and quarks, on which strong constraints exist from the measurements of the EDMs of the electron, neutron and mercury \cite{Brod:2013cka,Dekens:2013zca}. The strongest constraint arises from the electron EDM, for which the one-loop contribution arising from $\tilde\kappa_{\gamma Z}^{\rm eff}$ has been calculated in \cite{Koenig:2015pha}. The corresponding bounds are however model dependent, since they involve the couplings of first-generation fermions to the Higgs boson. If one assumes that these couplings are SM like, then the current experimental bound on the electron EDM \cite{Baron:2013eja} implies $|\tilde\kappa_{\gamma\gamma}+0.09\,\tilde\kappa_{\gamma Z}^{\rm eff}|<0.006$ at 90\% confidence level, but this bound can be avoided in models in which the Higgs boson does not couple to the electron. In order to be model independent we will not impose any EDM bound on $\tilde\kappa_{\gamma Z}^{\rm eff}$ in our analysis.

The structure of the results for vector mesons in (\ref{AVindir}) is interesting. The photon-pole diagram in Figure~\ref{fig:hMZ} yields contributions to the transverse form factors which are formally power-enhanced by $1/r_V=m_h^2/m_V^2$, and after squaring the form factors this enhancement more than compensates for the suppression factor $r_V$ in (\ref{hMZrates}). By power counting these are thus the leading contributions to the decay rates. However, these contributions are suppressed by $(\alpha/\pi)^2$, and hence there is an subtle interplay of suppression factors at work. We find that the photon-pole diagram gives the dominant contribution to the decay rates for light vector mesons, while it becomes subdominant for heavy vector mesons. This was also noted in \cite{Gao:2014xlv} but overlooked in \cite{Isidori:2013cla,Gonzalez-Alonso:2014rla}, where the photon-pole graph was neglected.

Contrary to the indirect contributions, which could be calculated in closed form, the direct contributions to the decay amplitudes can only be evaluated in a power series in $(\Lambda_{\rm QCD}/m_h)^2$ or $(m_q/m_h)^2$, where $\Lambda_{\rm QCD}$ is a hadronic scale and $m_q$ represents the masses of the constituent quarks of a given meson. The direct contributions to the $h\to MZ$ decay amplitudes with a pseudoscalar or longitudinally polarized vector meson in the final state arise from subleading-twist projections and hence are power suppressed. This is the main difference with regards to $h\to V\gamma$ decays, for which the direct contributions to the decay amplitudes arise at leading order \cite{Koenig:2015pha}. We discuss the detailed structure of these subleading-twist contributions in Appendix~\ref{App:SubleadingTwist}. For the purposes of illustration, we quote the result obtained for a pseudoscalar final-state meson $P$ in the limit where 3-particle LCDAs are neglected and where the asymptotic form $\phi_P(x)=6x(1-x)$ is used for the leading-twist LCDA. In this approximation, we find
\begin{equation}\label{FPZdirect}
   F_{\rm direct}^{PZ} 
   = \sum_q f_P^q\,a_q\,\kappa_q\,\frac{m_q}{m_h^2} \left( 2\mu_P - 3m_q \right) 
    \frac{1-r_Z^2+2r_Z\ln r_Z}{(1-r_Z)^3} \,,
\end{equation}
where the parameter $\mu_P=m_P^2/(m_{q_1}+m_{q_2})$ is related to the chiral condensate and governs the normalization of the twist-3 LCDAs.\footnote{Note that $\mu_\pi=m_\pi^2/(m_u+m_d)$ holds for charged and neutral pions, see e.g.\ \cite{Beneke:2002jn}.} 
This direct contribution is suppressed relative to the leading term in (\ref{AVindir}) by a factor $m_P^2/m_h^2$ or $m_q^2/m_h^2$, which makes it completely negligible. An analogous argument holds for the case of a longitudinally polarized vector meson. For the case of a transversely polarized vector meson the direct contribution arises from leading-twist projections. In the approximation where the asymptotic form $\phi_V^\perp(x)=6x(1-x)$ is used (the full expression is given in Appendix~\ref{App:SubleadingTwist}), we obtain \begin{equation}\label{Ftransdirect}
\begin{aligned}
   F_{\perp\,{\rm direct}}^{VZ} 
   &= \sum_q f_V^{q\perp} v_q\,\kappa_q\,\frac{3m_q}{2m_V}\,\frac{1-r_Z^2+2r_Z\ln r_Z}{(1-r_Z)^2} \,, \\
   \widetilde F_{\perp,{\rm direct}}^{VZ} 
   &= \sum_q f_V^{q\perp} v_q\,\tilde\kappa_q\,\frac{3m_q}{2m_V}\,\frac{1-r_Z^2+2r_Z\ln r_Z}{(1-r_Z)^2} \,,
\end{aligned}    
\end{equation}
which is parametrically of the same order as the indirect contribution given in (\ref{AVindir}). Numerically, the direct contributions are nevertheless strongly suppressed (see below). In the above expression $f_V^{q\perp}$ are the flavor-specific transverse decay constants of the meson, as defined in~\cite{Koenig:2015pha}.

Following \cite{Grossmann:2015lea}, we take $v=245.36$\,GeV for the Higgs vacuum expectation value at the electroweak scale and use $s_W^2=0.23126\pm 0.00005$ for the electroweak mixing angle. To obtain the $h\to MZ$ branching fraction we normalize the partial decay rates to the theoretical prediction for the total Higgs width in the SM, $\Gamma_h=(4.08\pm 0.16)$\,MeV, referring to the Higgs mass of $m_h=(125.09\pm 0.024)$\,GeV \cite{Heinemeyer:2013tqa}. 

\subsection{Hadronic input parameters}

The flavor-specific decay constants $f_M^q$ are the only hadronic quantities entering our predictions. We will assume that the heavy mesons $J/\psi$ and $\Upsilon(nS)$ can be described as pure $(c\bar c)$ and $(b\bar b)$ flavor states, and that the $\pi^0$ and $\rho^0$ mesons are pure $(u\bar u-d\bar d)/\sqrt2$ flavor states. We will furthermore assume unbroken isospin symmetry, such that $f_{\pi^0}^u=-f_{\pi^0}^d\equiv f_{\pi^0}/\sqrt2$ and analogously for $\rho^0$. For the mesons $\eta$ and $\eta'$ the contributions from up- and down-quark flavor states cancel out in the sum (in the isospin limit), and hence only the parameters $f_\eta^s$ and $f_{\eta'}^s$ are required. We adopt the FKS mixing scheme \cite{Feldmann:1998vh} and express these parameters as $f_\eta^s=-f_s\sin\varphi$ and $f_{\eta'}^s=f_s\cos\varphi$, where $f_s=(1.34\pm 0.06)\,f_{\pi^0}$ and $\varphi=(39.3\pm 1.0)^\circ$. This yields $f_\eta^s=-(110.7\pm 5.5)$\,MeV and $f_{\eta'}^s=(135.2\pm 6.4)$\,MeV, where the dominant errors are due to the uncertainty in the value of $f_s$.

For the vector mesons $\omega$ and $\phi$ we need the parameters $f_\omega^u=f_\omega^d$ and $f_\omega^s$, and analogously for $f_\phi^q$ and $f_\phi^s$. From measurements of the leptonic decay rates $V\to e^+ e^-$ one can determine the combinations \cite{Koenig:2015pha,Grossmann:2015lea}
\begin{equation}\label{fVvals}
\begin{aligned}
   f_\omega &= \frac{\sqrt2}{Q_u+Q_d} \sum_q f_\omega^q\,Q_q 
    = \sqrt2 \left( f_\omega^u - f_\omega^s \right) = (194.2\pm 2.1)\,\mbox{MeV} \,, \\
   f_\phi &= \frac{1}{Q_s} \sum_q f_\phi^q\,Q_q 
    = f_\phi^s - f_\phi^u = (223.0\pm 1.4)\,\mbox{MeV} \,.
\end{aligned}
\end{equation}
We shall adopt a simple flavor-mixing scheme for the $\omega-\phi$ system and express the physical mass eigenstates $|\omega\rangle$ and $|\phi\rangle$ in terms of the flavor eigenstates $|\omega_I\rangle=\frac{1}{\sqrt2}\left(|u\bar u\rangle+|d\bar d\rangle\right)$ and $|\phi_I\rangle=|s\bar s\rangle$ by means of the rotation by an angle $\theta$ (see \cite{Koenig:2015pha} for more details). In the limit where OZI-violating contributions are neglected, we can relate the matrix elements of the flavor-specific vector currents in (\ref{fPqdef}) to decay constants defined in terms of analogous matrix elements of the flavor eigenstates $|\omega_I\rangle$ and $|\phi_I\rangle$ with the corresponding flavor currents. Assuming isospin symmetry, this gives
\begin{equation}
   \sqrt2\,f_\omega^u = \cos\theta\,f_{\omega_I} \,, \qquad
   f_\omega^s = - \sin\theta\,f_{\phi_I} \,, \qquad 
   \sqrt2\,f_\phi^u = \sin\theta\,f_{\omega_I} \,, \qquad 
   f_\phi^s = \cos\theta\,f_{\phi_I} \,.
\end{equation}
It is now straightforward to solve relations (\ref{fVvals}) for $f_{\omega_I}$ and $f_{\phi_I}$ and express the flavor-specific decay constants in terms of the measured values $f_\omega$, $f_\phi$ and the mixing angle $\theta$. We obtain
\begin{equation}
   f_\omega^s = - \sin\theta \left( \cos\theta\,f_\phi + \frac{\sin\theta}{\sqrt2}\,f_\omega \right) , 
    \qquad
   f_\phi^s = \cos\theta \left( \cos\theta\,f_\phi + \frac{\sin\theta}{\sqrt2}\,f_\omega \right) .
\end{equation}
The corresponding expressions for $f_\omega^u$ and $f_\phi^u$ are readily obtained from (\ref{fVvals}). Existing estimates for the mixing angle $\theta$ derived from phenomenological analyses yield $\theta\approx 0.05$ \cite{Shifman:1978by} and $\theta\approx 0.06$ \cite{Benayoun:1999fv,Kucukarslan:2006wk}. In our analysis we use $\theta=0.06\pm 0.02$.

For the evaluation of the direct contributions to the transverse form factors in (\ref{Ftransdirect}) we also need the transverse decay constants $f_V^{q\perp}$ of vector mesons. Following \cite{Koenig:2015pha}, we compute them from the ratios $f_\rho^{q\perp}/f_\rho^q=0.72\pm 0.04$, $f_\omega^{q\perp}/f_\omega^q=0.71\pm 0.05$, $f_\phi^{q\perp}/f_\phi^q=0.76\pm 0.04$ for light mesons, and $f_{J/\psi}^\perp/f_{J/\psi}=0.91\pm 0.14$, $f_{\Upsilon(1S)}^\perp/f_{\Upsilon(1S)}=1.09\pm 0.02$, $f_{\Upsilon(2S)}^\perp/f_{\Upsilon(2S)}=1.08\pm 0.02$, $f_{\Upsilon(3S)}^\perp/f_{\Upsilon(3S)}=1.07\pm 0.03$ for heavy mesons, where the scale-dependent transverse decay constants refer to the scale $\mu=2$\,GeV. We evolve these quantities to the hard scale $\mu_{hZ}\approx 58.6$\,GeV using two-loop renormalization-group equations \cite{Koenig:2015pha}.

\subsection{Structure of the form factors and sensitivity to new physics}

We briefly explore the structure of the form factors for a few representative cases, using the central values for the decay constants. For the pseudoscalar mesons we find (units are MeV)
\begin{equation}
   F^{\pi^0 Z}\approx 46.1\,\kappa_Z \,, \qquad
   F^{\eta Z}\approx 27.7\,\kappa_Z \,, \qquad
   F^{\eta' Z}\approx - 33.8\,\kappa_Z \,.
\end{equation}
The direct contributions to these form factors are extremely small. For the pion case, e.g., they yield a relative correction factor $(1+2.7\cdot 10^{-7}\kappa_u+5.9\cdot 10^{-7}\kappa_d)$. The results for vector mesons have a richer structure, since the loop-induced photon-pole contributions involve several new-physics parameters. We obtain (units are again MeV)
\begin{equation}
\begin{aligned}
   F_\parallel^{\rho^0 Z}
   &\approx 41.11\,\kappa_Z - 0.98 + 0.41\,\kappa_{\gamma Z}^{\rm eff} \,, \\
   F_\perp^{\rho^0 Z}
   &\approx - 2640 + 1102\,\kappa_{\gamma Z}^{\rm eff} + 41.11\,\kappa_Z  
    + 0.018\,\kappa_d + 0.005\,\kappa_u \,, \\
   F_\parallel^{\omega Z}
   &\approx - 7.14\,\kappa_Z - 0.29 + 0.12\,\kappa_{\gamma Z}^{\rm eff} \,, \\
   F_\perp^{\omega Z}
   &\approx - 775.4 + 323.7\,\kappa^{\rm eff}_{\gamma Z} - 7.14\,\kappa_Z 
    + 0.032\,\kappa_s - 0.014\,\kappa_d + 0.004\,\kappa_u \,, \\
   F_\parallel^{\phi Z}
   &\approx - 40.41\,\kappa_Z + 0.48 - 0.20\,\kappa_{\gamma Z}^{\rm eff} \,, \\
   F_\perp^{\phi Z}
   &\approx 744.1 - 310.7\,\kappa_{\gamma Z}^{\rm eff} - 40.41\,\kappa_Z 
    - 0.43\,\kappa_s - 0.0007\,\kappa_d + 0.0002\,\kappa_u \,, \\
   F_\parallel^{J/\psi\,Z}
   &\approx 38.69\,\kappa_Z - 1.75 + 0.73\,\kappa_{\gamma Z}^{\rm eff} \,, \\
   F_\perp^{J/\psi\,Z}
   &\approx - 294.8 + 123.1\,\kappa_{\gamma Z}^{\rm eff} + 38.69\,\kappa_Z  
    + 1.95\,\kappa_c \,, \\
   F_\parallel^{\Upsilon(1S)\,Z}
   &\approx - 119.63\,\kappa_Z + 1.52 - 0.64\,\kappa_{\gamma Z}^{\rm eff} \,, \\
   F_\perp^{\Upsilon(1S)\,Z}
   &\approx 26.83 - 11.20\,\kappa_{\gamma Z}^{\rm eff} - 119.62\,\kappa_Z 
    - 10.47\,\kappa_b \,.
\end{aligned}
\end{equation}
Similar expressions are obtained for the other $\Upsilon(nS)$ states. The CP-odd transverse form factors $\widetilde F_\perp^{VZ}$ are given by similar expressions as $F_\perp^{VZ}$, but with the constant terms omitted and with the replacements $\kappa_Z\to 0$ and $\kappa_{\gamma Z}^{\rm eff}\to\tilde\kappa_{\gamma Z}^{\rm eff}$, $\kappa_q\to\tilde\kappa_q$. For instance,
\begin{equation}
   \widetilde F_\perp^{\rho^0 Z}
   \approx 1102\,\tilde\kappa_{\gamma Z}^{\rm eff}  
    + 0.018\,\tilde\kappa_d + 0.005\,\tilde\kappa_u \,.
\end{equation}

In the above expressions the terms proportional to $\kappa_Z$ are the indirect contributions involving the $hZZ$ vertex (third graph in Figure~\ref{fig:hMZ}), while the constant terms and the pieces proportional to $\kappa_{\gamma Z}^{\rm eff}$ ($\tilde\kappa_{\gamma Z}^{\rm eff}$) are the indirect contributions involving the effective $h\gamma Z$ vertex (fourth graph). The terms involving the $\kappa_q$ ($\tilde\kappa_q$) parameters of the quarks contained in the meson $V$ are the direct contributions from (\ref{Ftransdirect}). With the exception of the bottomonium states, we observe that the transverse form factors are much larger than the longitudinal ones, an effect that results from the photon-pole contribution and is most pronounced for the lightest mesons. The enhancement of the transverse form factors is sufficiently large to overcome the phase-space suppression in front of these form factors in (\ref{hMZrates}). It follows that the $h\to VZ$ decay rates are sensitive to the new-physics coefficients $\kappa_{\gamma Z}^{\rm eff}$ and $\tilde\kappa_{\gamma Z}^{\rm eff}$. This will be studied in more detail in Section~\ref{sec:hVZbsm}. On the other hand, the sensitivity to the Yukawa coupling of the light quarks (parameters $\kappa_q$ and $\tilde\kappa_q$), which is induced by the direct contributions in (\ref{Ftransdirect}), is too weak to be of any relevance. It would be a good approximation to neglect these direct contributions altogether. We will instead keep them at their SM values.

\subsection{SM branching ratios}

In Table~\ref{tab:htoMZ} we show our predictions for the branching fractions of several $h\to MZ$ decay modes. We show the dominant theoretical uncertainties, which arise from the uncertainties in the meson decay constants and the theoretical estimate for the total width of the Higgs boson. The relevant decay constants are compiled in the last column of the table. In the case of $\eta$ and $\eta'$ mesons in the final state we neglect the loop-suppressed contributions from the two-gluon LCDA of mesons with a flavor-singlet component. For the related case of $Z\to\eta^{(\prime)}\gamma$ decays these effects were studied in \cite{Alte:2015dpo} and found to be very small. The branching ratios range from $6\cdot 10^{-7}$ for the decay $h\to\omega Z$ up to $1.5\cdot 10^{-5}$ for the decay $h\to \Upsilon(1S)\,Z$. 

\begin{table}
\centering
\renewcommand{\arraystretch}{1.3}
\begin{tabular}{|c|c|c|}
\hline
Decay mode & Branching ratio [$10^{-6}$] & Decay constant [MeV] \\
\hline
$h\to\pi^0 Z$ & $2.30\pm 0.01_f\pm 0.09_{\Gamma_h}$ & $130.4\pm 0.2$ \\
$h\to\eta Z$ & $0.83\pm 0.08_f\pm 0.03_{\Gamma_h}$ & $f_\eta^s=-110.7\pm 5.5$ \\
$h\to\eta' Z$ & $1.24\pm 0.12_f \pm 0.05_{\Gamma_h}$ & $f_{\eta'}^s=135.2\pm 6.4$ \\
$h\to\rho^0 Z$ & $7.19\pm 0.09_f\pm 0.28_{\Gamma_h}$ & $216.3\pm 1.3$ \\
$h\to\omega Z$ & $0.56\pm 0.01_f\pm 0.02_{\Gamma_h}$ & $f_\omega=194.2\pm 2.1\,,$
 ~ $f_\omega^s=-13.8\pm 4.8$ \\
$h\to\phi Z$ & $2.42\pm 0.05_f\pm 0.09_{\Gamma_h}$ & $f_\phi=223.0\pm 1.4\,,$
 ~ $f_\phi^s=230.4\pm 2.6$ \hspace{0mm} \\
$h\to J/\psi\,Z$ & $2.30\pm 0.06_f\pm 0.09_{\Gamma_h}$ & $403.3\pm 5.1$ \\
$h\to\Upsilon(1S)\,Z$ & $15.38\pm 0.21_f\pm 0.60_{\Gamma_h}$ & $684.4\pm 4.6$ \\
$h\to\Upsilon(2S)\,Z$ & $7.50\pm 0.14_f\pm 0.29_{\Gamma_h}$ & $475.8\pm 4.3$ \\
$h\to\Upsilon(3S)\,Z$ & $5.63\pm 0.10_f\pm 0.22_{\Gamma_h}$ & $411.3\pm 3.7$ \\
\hline 
\end{tabular}
\caption{\label{tab:htoMZ}
SM predictions for the branching ratios of the rare exclusive decays $h\to MZ$ for a variety of pseudoscalar and vector mesons. The decay rates are normalized to the SM prediction for the total Higgs width. The quoted errors show the uncertainties related to the decay constants and the total width.}
\end{table} 

Let us briefly compare our results with previous computations in the literature, which use almost identical values for the hadronic input parameters. The authors of \cite{Isidori:2013cla} reported the branching ratios (all in units of $10^{-6}$)  $\mbox{Br}(\pi^0 Z)=3.0$, $\mbox{Br}(\phi Z)=2.2$, $\mbox{Br}(\rho^0 Z)=1.2$ and $\mbox{Br}(J/\psi\,Z)=2.2$, while the authors of \cite{Gonzalez-Alonso:2014rla} obtained $\mbox{Br}(J/\psi\,Z)=1.7$ and $\mbox{Br}(\Upsilon(1S)\,Z)=16$. In these papers the indirect contributions to the $h\to VZ$ modes involving the $h\gamma Z$ vertex have been neglected. As a result, the rate for $h\to\rho^0 Z$ decay in particular comes out much too small. In \cite{Gao:2014xlv}, the branching ratios $\mbox{Br}(J/\psi\,Z)=3.2$ and $\mbox{Br}(\Upsilon(1S)\,Z)=17$ were presented, whereas the authors of \cite{Modak:2014ywa} found $\mbox{Br}(J/\psi\,Z)=3.6$ and $\mbox{Br}(\Upsilon(1S)\,Z)=22$. The interference terms involving the two indirect contributions in these works have the opposite sign compared to our findings, and hence the branching fractions come out too high.

The strong suppression of the direct contributions, which contain all sensitivity to the quark Yukawa couplings, makes the $h\to MZ$ decay modes unsuitable for searches for new-physics effects on the Yukawa couplings of the light quarks. Instead, the pseudoscalar modes could serve as ``standard candles'', since the calculation of their decay rates yields highly accurate, model-independent predictions, subject to electroweak corrections only. Non-standard effects only enter via the Higgs coupling to $Z$ bosons (as parameterized by $\kappa_Z$), which is constrained to be close to~1 by phenomenological analyses of the LHC data \cite{Khachatryan:2016vau}. The modes with vector mesons are sensitive to new-physics effects in the effective $h\gamma Z$ vertex. This will be explored in the next section.

\subsection{Sensitivity to new physics}
\label{sec:hVZbsm}

\begin{figure}[t]
\begin{center}
\raisebox{0pt}{\includegraphics[width=0.46\textwidth]{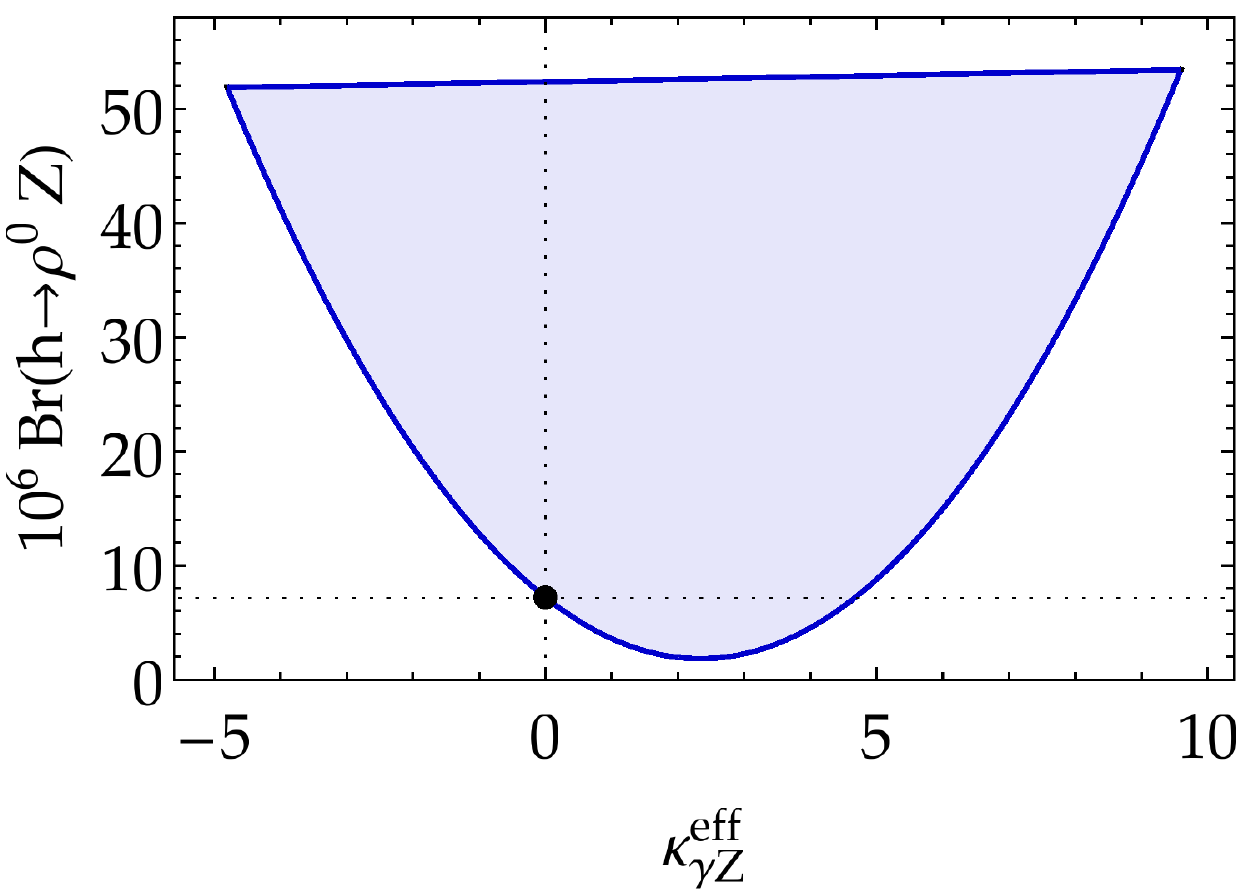}} \quad
\raisebox{0pt}{\includegraphics[width=0.46\textwidth]{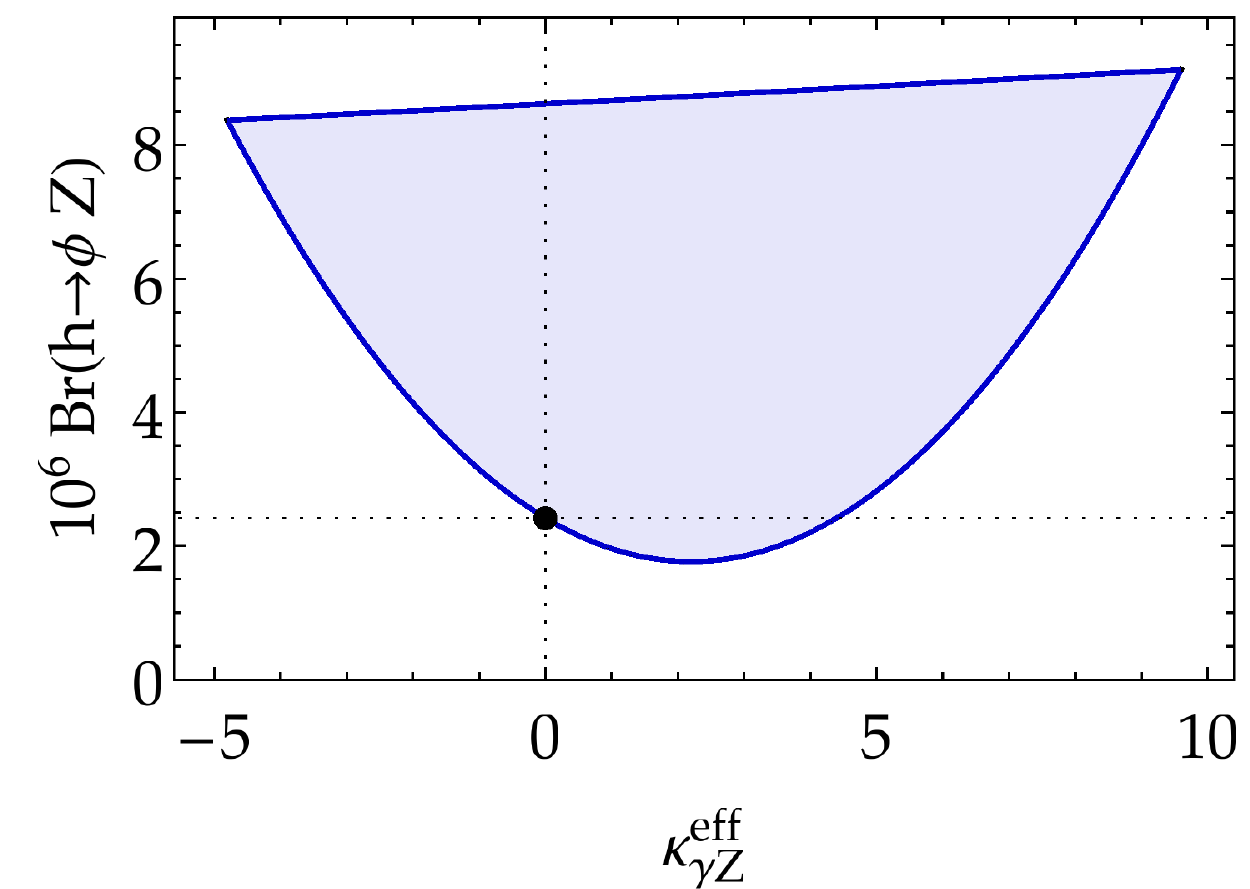}}
\raisebox{0pt}{\includegraphics[width=0.46\textwidth]{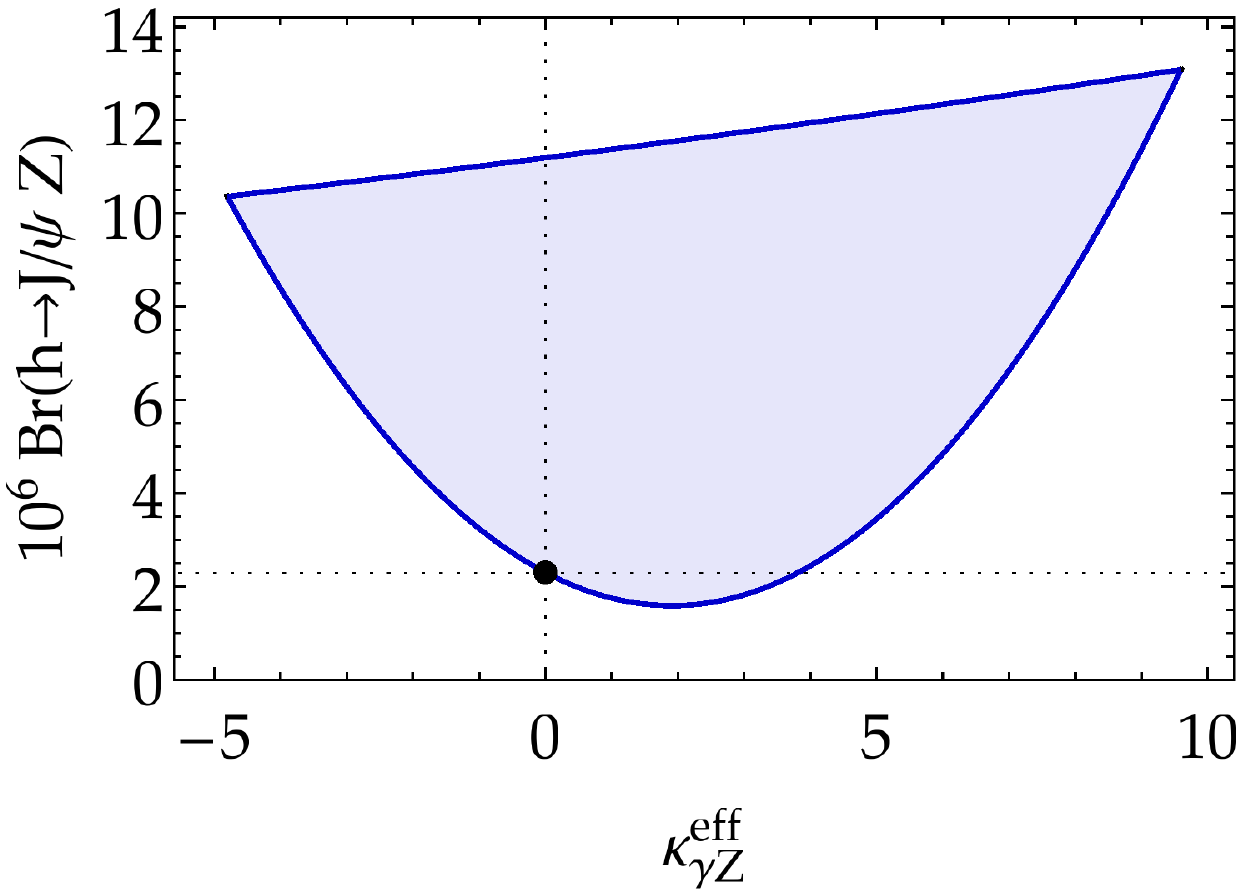}} \quad
\raisebox{0pt}{\includegraphics[width=0.46\textwidth]{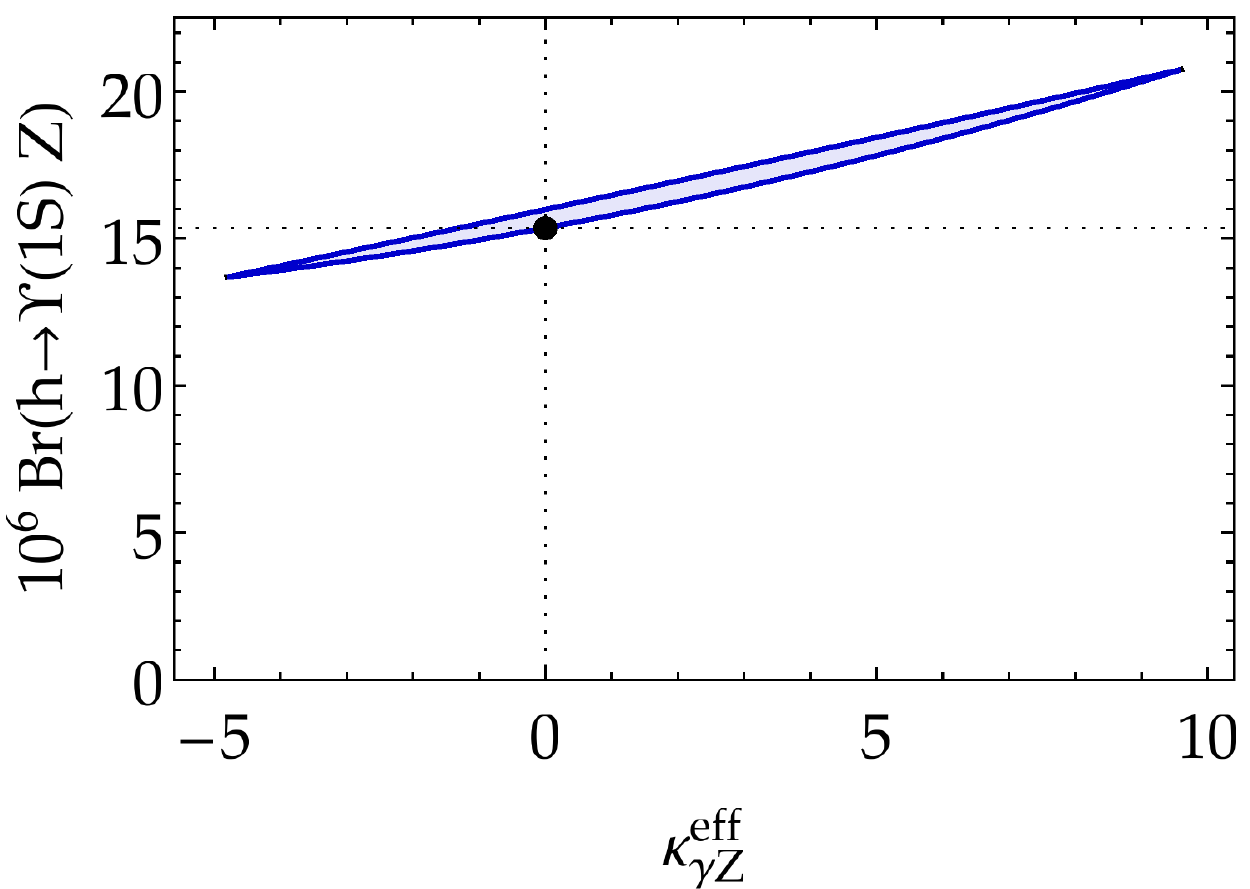}}
\parbox{15.5cm}
{\caption{\label{fig:hVZbsm}
Allowed regions for four of the $h\to VZ$ branching ratios in the presence of new-physics contributions to the effective $h\gamma Z$ vertices. The parameters $\kappa_{\gamma Z}^{\rm eff}$ and $\tilde\kappa_{\gamma Z}^{\rm eff}$ are varied in the range allowed by the constraint (\ref{gaZrange}) derived from the $h\to\gamma Z$ decay rate. The black dots show the SM values.}}
\end{center}
\end{figure}

In Figure~\ref{fig:hVZbsm} we show four of the $h\to VZ$ branching ratios as functions of the parameters $\kappa_{\gamma Z}^{\rm eff}$ and $\tilde\kappa_{\gamma Z}^{\rm eff}$ defined in (\ref{keffdef}), which parameterize possible new-physics contributions to the effective $h\gamma Z$ vertices. We vary these parameters within the range allowed by the constraint (\ref{gaZrange}). The lower, parabola-shaped boundaries of the shaded regions correspond to $\tilde\kappa_{\gamma Z}^{\rm eff}=0$, while the upper, straight-line boundaries are obtained when $|\tilde\kappa_{\gamma Z}^{\rm eff}|$ takes the maximum value allowed for a given value of $\kappa_{\gamma Z}^{\rm eff}$. We only show the central values of the branching ratios. In all cases the parametric uncertainties are below the 5\% level, see Table~\ref{tab:htoMZ}. We observe that in the presence of new physics the $h\to VZ$ branching ratios can be significantly enhanced (or slightly reduced) compared with their SM values indicated by the black dots. The allowed ranges are shown in Table~\ref{tab:htoMZbsm}. By the time the rare exclusive decays $h\to VZ$ can be explored experimentally, it is likely that the $h\to\gamma Z$ rate will have been measured with high accuracy. As is evident from (\ref{gaZrange}), this will constrain the new-physics parameters to lie on a circle centered at $\kappa_{\gamma Z}^{\rm eff}=2.395$ and $\tilde\kappa_{\gamma Z}^{\rm eff}=0$. A measurement of some of the $h\to VZ$ decay rates could help to lift some of the degeneracies and determine $\kappa_{\gamma Z}^{\rm eff}$ and $|\tilde\kappa_{\gamma Z}^{\rm eff}|$ individually. For example, the $h\to\Upsilon(1S)\,Z$ branching ratio directly probes the value of $\kappa_{\gamma Z}^{\rm eff}$.

\begin{table}[t]
\centering
\renewcommand{\arraystretch}{1.3}
\begin{tabular}{|c|c|c|}
\hline
Decay mode & SM branching ratio [$10^{-6}$] & Range with new physics [$10^{-6}$] \\
\hline
$h\to\rho^0 Z$ & $7.19\pm 0.29$ & $1.83 - 53.3$ \\
$h\to\omega Z$ & $0.56\pm 0.02$ & $0.06 - 4.56$ \\
$h\to\phi Z$ & $2.42\pm 0.10$ & $1.77 - 9.12$ \\
$h\to J/\psi\,Z$ & $2.30\pm 0.11$ & $1.59 - 13.1$ \\
$h\to\Upsilon(1S)\,Z$ & $15.38\pm 0.64$ & $13.7 - 20.8$ \\
\hline 
\end{tabular}
\caption{\label{tab:htoMZbsm}
Allowed ranges for the $h\to VZ$ branching ratios in the presence of new-physics contributions to the effective $h\gamma Z$ vertices. Only central values are shown.}
\end{table} 

Let us briefly comment on the prospects for probing flavor-changing Higgs couplings with $h\to MZ$ decays. Then only the direct contributions to the form factors arise at leading order in $\alpha_\mathrm{EW}$. The leading decay rates are then the ones with transversely polarized vector mesons. For a final-state vector meson containing the quark flavors $q$ and $q'$, we obtain using the asymptotic form of the leading-twist LCDA
\begin{equation}\label{eq22}
   \Gamma(h\to V_{qq'} Z) 
   = \frac{9m_h (f_V^\perp)^2}{8\pi v^2}\,v_q^2 \left( |Y_{qq'}|^2 + |Y_{q'q}|^2 \right) 
    \frac{r_Z}{(1-r_Z)^3} \left( 1 - r_Z^2 + 2r_Z\ln r_Z \right)^2 .
\end{equation}
The complete expression is given in Appendix~\ref{App:SubleadingTwist}. The weakest indirect bounds on flavor-changing Higgs couplings refer to possible $hb\bar s$ interactions, for which the bounds derived from $B_s\!-\!\bar B_s$ mixing imply $|Y_{bs}|^2+|Y_{sb}|^2<7\cdot 10^{-6}$ (at 95\% confidence level) \cite{Harnik:2012pb}. Using a typical value $f_{B_s^*}^\perp\approx 0.2$\,GeV for the transverse decay constant of the $B_s^*$ meson, we estimate that the $h\to B_s^* Z$ branching fraction is bounded by
\begin{equation}
   \mbox{Br}(h\to B_s^* Z)\approx 2.3\cdot 10^{-11}\times\frac{|Y_{bs}|^2+|Y_{sb}|^2}{10^{-5}} \,.
\end{equation}
Detecting such a small branching fraction seems unimaginable at any currently envisaged particle collider. Note that one-loop electroweak corrections in the SM can also give rise to flavor off-diagonal $h\to MZ$ decays, where $M$ can be either a pseudoscalar or a vector meson. The corresponding contributions to the $h\to B_s^{(*)} Z$ form factors are of order $F^{B_s^{(*)}Z}\sim\frac{\alpha}{4\pi s_W^2}\,|V_{tb} V_{ts}^*|\sim 10^{-4}$, yielding tiny branching ratios of order $4\cdot 10^{-13}$.

\section{\boldmath Weak radiative hadronic decays $h\to M^+ W^-$}
\label{sec:hMW}

\begin{figure}
\begin{center}
\raisebox{0pt}{\includegraphics[width=0.25\textwidth]{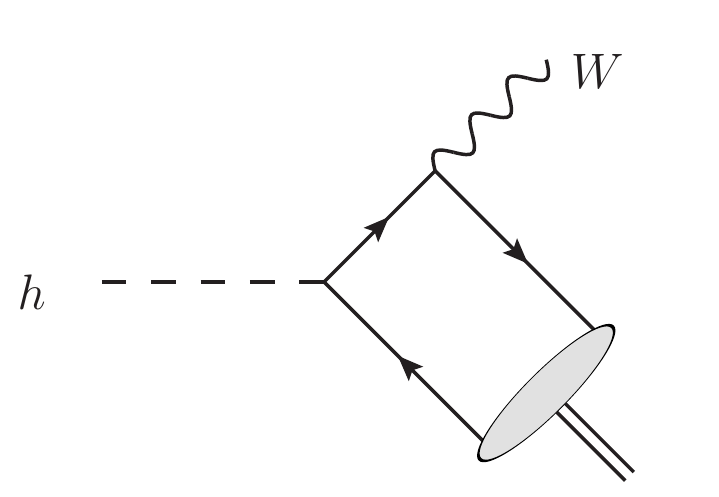}}
\raisebox{-5pt}{\includegraphics[width=0.25\textwidth]{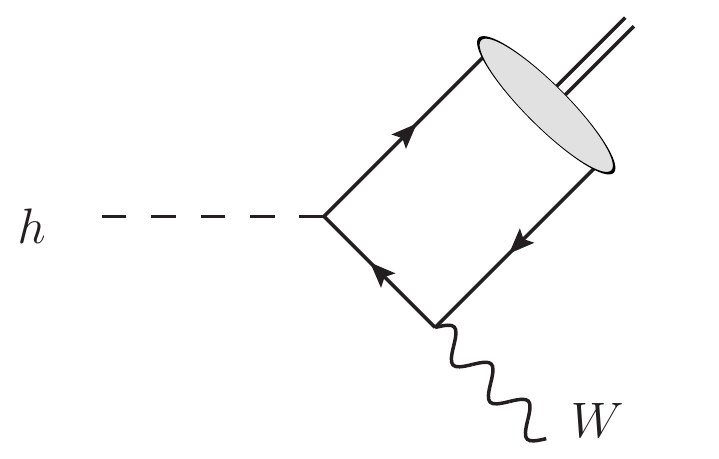}}
\raisebox{2pt}{\includegraphics[width=0.235\textwidth]{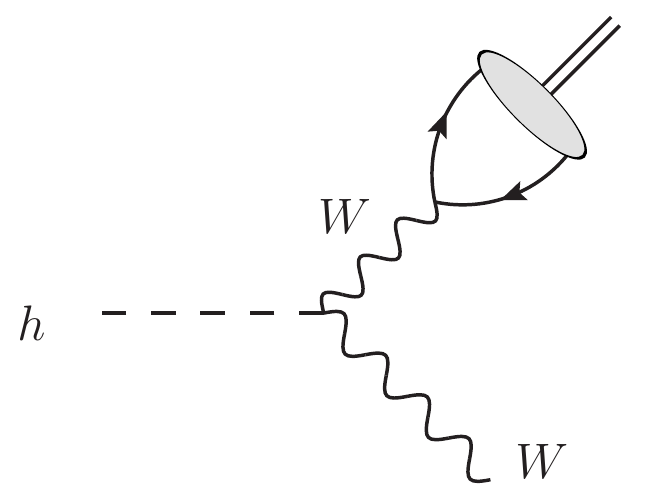}}
\parbox{15.5cm}
{\caption{\label{fig:hMW}
Leading-order diagrams contributing to the decays $h\to M^+ W^-$.}}
\end{center}
\end{figure}

The weak radiative Higgs decays into final states containing a $W$ boson are in many ways similar to the $h\to MZ$ decays just discussed. However, since the charged-current interactions in the SM are flavor changing, the final-state meson $M\sim(u_i\bar d_j)$ is flavor non-diagonal and its production involves the corresponding CKM matrix element $V_{ij}$. We show the relevant Feynman diagrams in Figure~\ref{fig:hMW}. As we will show, an interesting probe of flavor-changing Higgs couplings involving the third-generation fermions arises when the virtual quark in the direct amplitude is a top quark and the indirect amplitude is CKM suppressed. Before we discuss this case, we focus on a scenario where the Higgs-boson couplings are flavor diagonal.

\subsection{Decay rates in the case of flavor-diagonal Higgs couplings}

\begin{table}
\centering
\renewcommand{\arraystretch}{1.3}
\begin{tabular}{|c|c|c|}
\hline
Decay mode & Branching ratio [$10^{-6}$] & Decay constant [MeV] \\
\hline
$h\to\pi^+ W^-$ & $4.30\pm 0.01_f\pm 0.00_\mathrm{CKM}\pm 0.17_{\Gamma_h}$ & $130.4\pm 0.2$ \\
$h\to\rho^+ W^-$ & $10.92\pm 0.15_f\pm 0.00_\mathrm{CKM}\pm 0.43_{\Gamma_h}$ & $207.8\pm 1.4$ \\ 
$h\to K^+ W^-$ & $0.33\pm 0.00_f\pm 0.00_\mathrm{CKM}\pm 0.01_{\Gamma_h}$ & $156.2\pm 0.7$ \\
$h\to K^{*+} W^-$ & $0.56\pm 0.03_f\pm 0.00_\mathrm{CKM}\pm 0.02_{\Gamma_h}$ & $203.2\pm 5.9$ \\
$h\to D^+ W^-$ & $0.56\pm 0.03_f\pm 0.04_\mathrm{CKM}\pm 0.02_{\Gamma_h}$ & $204.6\pm 5.0$ \\
$h\to D^{*+} W^-$ & $1.04\pm 0.12_f\pm 0.07_\mathrm{CKM}\pm 0.04_{\Gamma_h}$ & $278\pm 16$ \\
$h\to D_s^+ W^-$ & $17.12\pm 0.61_f \pm 0.56_\mathrm{CKM} \pm 0.67_{\Gamma_h}$ & $257.5\pm 4.6$ \\
$h\to D_s^{*+} W^-$ & $25.10\pm 1.45_f\pm 0.81_\mathrm{CKM}\pm 0.98_{\Gamma_h}$ & $311\pm 9$ \\
$h\to B^+ W^-$ & $(1.54\pm 0.15_f\pm 0.36_\mathrm{CKM}\pm 0.06_{\Gamma_h})\cdot 10^{-4}$ & $186\pm 9$ \\
$h\to B^{*+} W^-$ &$(1.41\pm 0.10_f\pm 0.34_\mathrm{CKM}\pm 0.06_{\Gamma_h})\cdot 10^{-4}$ & $175\pm 6$ \\
$h\to B^+_c W^-$ &$(8.21\pm 0.57_f\pm 0.52_\mathrm{CKM}\pm 0.32_{\Gamma_h})\cdot 10^{-2}$ & $434\pm 15$ \\
\hline 
\end{tabular}
\caption{SM predictions for the branching ratios of the rare exclusive decays $h\to M^+W^-$ for a variety of pseudoscalar and vector mesons. The decay rates are normalized to the SM prediction for the total Higgs width. The quoted errors show the uncertainties related to the decay constants, the relevant CKM matrix elements and the total width.}
\label{tab:WSMBR}
\end{table} 

In analogy with (\ref{hPZampl}), the most general parameterization of the decay amplitudes into pseudoscalar mesons can be written as (again with $M=P,V_\parallel$) 
\begin{equation}\label{hPWampl}
   i{\cal A}(h\to P^+ W^-)
   = \frac{g}{\sqrt2 v}\,k\cdot\varepsilon_W^*\,F^{MW} \,,
\end{equation}
where $k$ is the meson momentum. The $W$ boson is longitudinally polarized in this case. The most general parameterization of the decay amplitudes into vector mesons can be chosen as
\begin{equation}\label{hVperpWampl}
   i{\cal A}(h\to V^+ W^-) 
   = - \frac{g m_V}{\sqrt2 v}
    \left[ \varepsilon_V^{\parallel *}\cdot\varepsilon_W^{\parallel *}\,F_\parallel^{VW} 
    + \varepsilon_V^{\perp *}\cdot\varepsilon_W^{\perp *}\,F_\perp^{VW} 
    + \frac{\epsilon_{\mu\nu\alpha\beta}\,k^\mu q^\nu\varepsilon_V^{*\alpha}\varepsilon_W^{*\beta}}%
           {\left[ (k\cdot q)^2-k^2 q^2\right]^{1/2}}\,\widetilde F_\perp^{VW} \right] ,
\end{equation}
where $q$ denotes the momentum of the $W$ boson, and the longitudinal and transverse polarization vectors have been defined in (\ref{hVpolarizations}). The total decay rates are found to be (defining $r_W=m_W^2/m_h^2$)
\begin{equation}\label{eq:hMWBR}
\begin{aligned}
   \Gamma(h\to P^+ W^-) 
   &= \frac{m_h^3}{32\pi v^4}\,\lambda^{3/2}(1,r_W,r_P) \left| F^{MW} \right|^2 , \\
   \Gamma(h\to V^+ W^-) 
   &= \frac{m_h^3}{32\pi v^4}\,\lambda^{1/2}(1,r_W,r_V)\,(1-r_W-r_V)^2 \\
   &\quad\times \left[
    \big| F_\parallel^{VW} \big|^2 + \frac{8r_V r_W}{(1-r_W-r_V)^2}
    \Big( \big| F_\perp^{VW} \big|^2 + \big| \widetilde F_\perp^{VW} \big|^2 \Big) \right] .
\end{aligned}
\end{equation}

In close similarity with (\ref{AVindir}), the indirect contributions to the form factors arising from the last diagram in Figure~\ref{fig:hMW} are found to be
\begin{equation}\label{APindirW}
   F_{\rm indirect}^{PW} = \kappa_W f_P V_{ij} \,, \qquad
   F_{\parallel\,\rm indirect}^{VW} = F_{\perp\,\rm indirect}^{VW}
    = \frac{\kappa_W f_V V_{ij}}{1-r_V/r_W} \,, \qquad
   \widetilde F_{\perp\,\rm indirect}^{VW} = 0 \,,
\end{equation}
where $V_{ij}$ is the relevant CKM matrix element. The direct contributions to the form factors are once again power suppressed and can be neglected to an excellent approximation. In this limit, we obtain for the decay rates 
\begin{equation*}
   \Gamma(h\to P^+ W^-) 
   = \kappa_W^2 \left| V_{ij} \right|^2 \frac{m_h^3\,f_P^2}{32\pi v^4}\,
    \lambda^{3/2}(1,r_W,r_P) \,, \hspace{4.5cm}
\end{equation*}
\begin{equation}\label{eq:hMWBR_approx}
   \Gamma(h\to V^+ W^-) 
   = \kappa_W^2 \left| V_{ij} \right|^2 \frac{m_h^3\,f_V^2}{32\pi v^4}\,
    \frac{\lambda^{1/2}(1,r_W,r_V)}{\left(1-r_V/r_W\right)^2} 
    \Big[ \lambda(1,r_W,r_V) + 12 r_V r_W \Big] \,.
\end{equation}
In Table~\ref{tab:WSMBR} we present our numerical predictions for the $h\to M^+ W^-$ branching ratios obtained by normalizing the partial decay rates to the total width of the Higgs boson. We set $\kappa_W=1$, noting that all rates are proportional to $\kappa_W^2$. The last column shows the relevant values of the decay constant. For the decay constants of the heavy mesons $D_{(s)}^*$, $B^*$ and $B_c$ we adopt the values obtained from two recent lattice calculations \cite{Becirevic:2012ti,Colquhoun:2015oha}, while all the other decay constants are taken from the recent compilation in \cite{Grossmann:2015lea}. We can again compare our results to the ones obtained in \cite{Isidori:2013cla}, which are (in units of $10^{-6}$) $\mbox{Br}(\pi^+ W^-)=6$, $\mbox{Br}(\rho^+ W^-)=8$, $\mbox{Br}(K^+ W^-)=0.4$, $\mbox{Br}(D^+ W^-)=0.7$, $\mbox{Br}(D^{*+} W^-)=1.2$, $\mbox{Br}(D_s^+ W^-)=21$ and $\mbox{Br}(D_s^{*+} W^-)=35$. We have been unable to trace the origins of the slight numerical differences with our results. 

\subsection{Effects of non-standard, flavor-changing Higgs couplings}

The situation changes when flavor-changing Higgs couplings, which are absent in the SM, are taken into account. Then the power-suppressed direct contributions can be enhanced by a factor of $m_t$, if a top quark is propagating between the Higgs and $W$ vertices in the first two graphs in Figure~\ref{fig:hMW}. Also, these contributions come with different CKM factors than the indirect ones. In cases where $|V_{ij}|\ll 1$, these two effects can compensate (at least to some extent) for the power suppression of the direct contribution. For a pseudoscalar or longitudinally polarized vector meson in the final state, we find (neglecting terms not enhanced by the top-quark mass)
\begin{equation}\label{eq:FPW-}
   F_{\rm direct}^{MW} = \frac{v}{2\sqrt2}\,\frac{f_M\,m_t}{m_h^2}\,Y_{it}\,V_{tj}
    \int_0^1\!dx\,\frac{\phi_M(x)}{r_t-\bar x-r_W x} \,,
\end{equation}
where $r_t=m_t^2/m_h^2$, $\phi_M(x)$ is the leading-twist LCDA of the meson $M$, and $\bar x\equiv 1-x$. For the case of a transversely polarized vector meson, we find instead
\begin{equation}\label{eq30}
\begin{aligned}
   F_{\perp\,\rm direct}^{V W} 
   &= - \frac{v}{4\sqrt2}\,\frac{f_V^\perp}{m_V}\,(1-r_W)
    \int_0^1\!dx\,\phi_V^\perp(x) \sum_k
    \left( \frac{\left(Y_u\right)_{ki}^* V_{kj}}{r_t\,\delta_{k3}-\bar x-r_W x}
    - \frac{V_{ik} \left(Y_d\right)_{kj}}{x+r_W\bar x} \right) \\
   &\approx - \frac{v}{4\sqrt2}\,\frac{f_V^\perp}{m_V}\,Y_{ti}^*\,V_{tj}\,(1-r_W)
    \int_0^1\!dx\,\frac{\phi_V^\perp(x)}{r_t-\bar x-r_W x} \,, \\
   \widetilde F_{\perp\,\rm direct}^{V W} 
   &= -i\,\frac{v}{4\sqrt2}\,\frac{f_V^\perp}{m_V}\,(1-r_W)
    \int_0^1\!dx\,\phi_V^\perp(x) \sum_k
    \left( \frac{\left(Y_u\right)_{ki}^* V_{kj}}{r_t\,\delta_{k3}-\bar x-r_W x}
    + \frac{V_{ik} \left(Y_d\right)_{kj}}{x+r_W\bar x} \right) \\
   &\approx -i\,\frac{v}{4\sqrt2}\,\frac{f_V^\perp}{m_V}\,Y_{ti}^*\,V_{tj}\,(1-r_W)
    \int_0^1\!dx\,\frac{\phi_V^\perp(x)}{r_t-\bar x-r_W x} \,,
\end{aligned}
\end{equation}
where in the last step we have used that only the Yukawa coupling involving the top quark can be sufficiently large to make this contribution relevant, see Section~\ref{sec:intro}. In this case there is no enhancement by the top-quark mass; however, the enhancement factor $v/m_V$ relative to (\ref{eq:FPW-}) potentially renders the rates for decays into transversely polarized vector mesons of a similar magnitude as those into longitudinally polarized ones. 

\begin{figure}
\begin{center}
\includegraphics[width=0.46\textwidth]{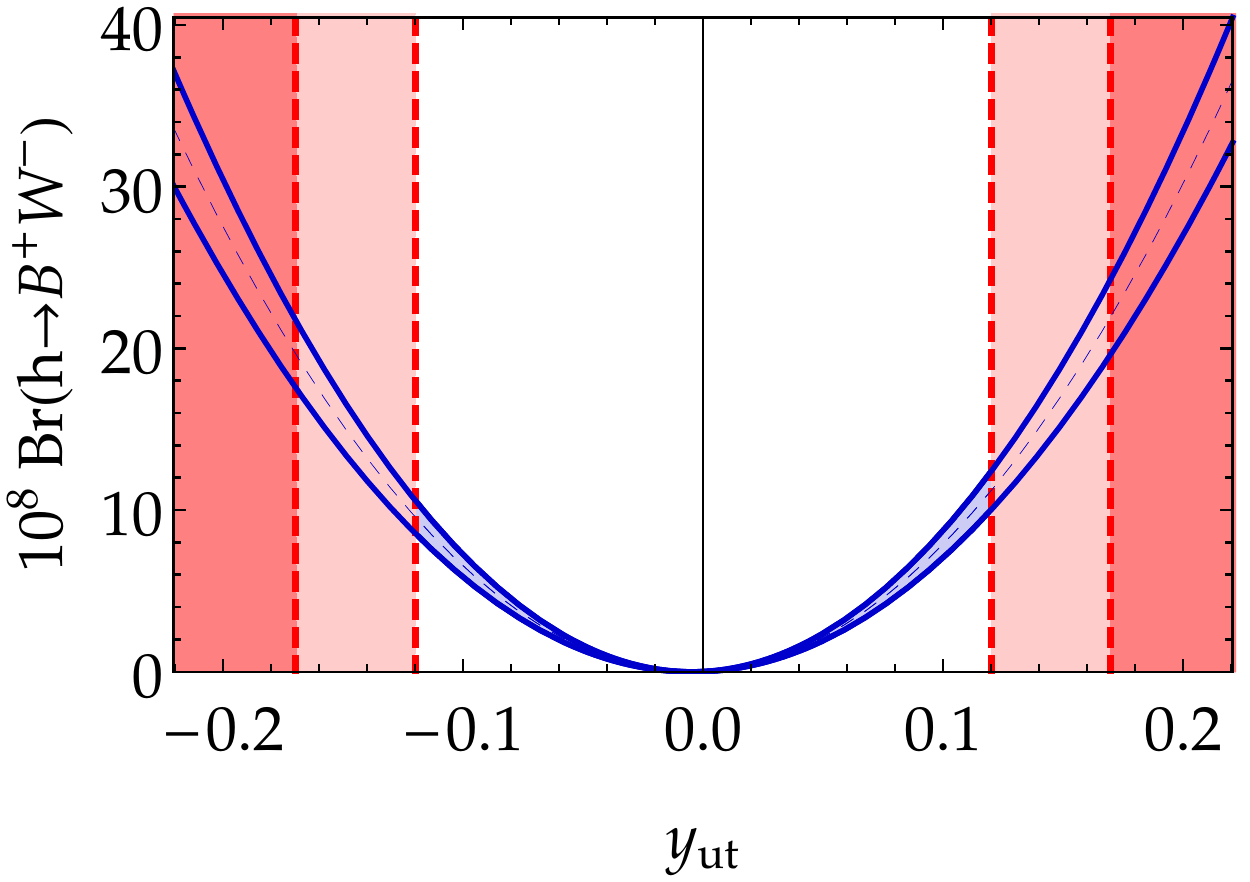}
\quad
\includegraphics[width=0.46\textwidth]{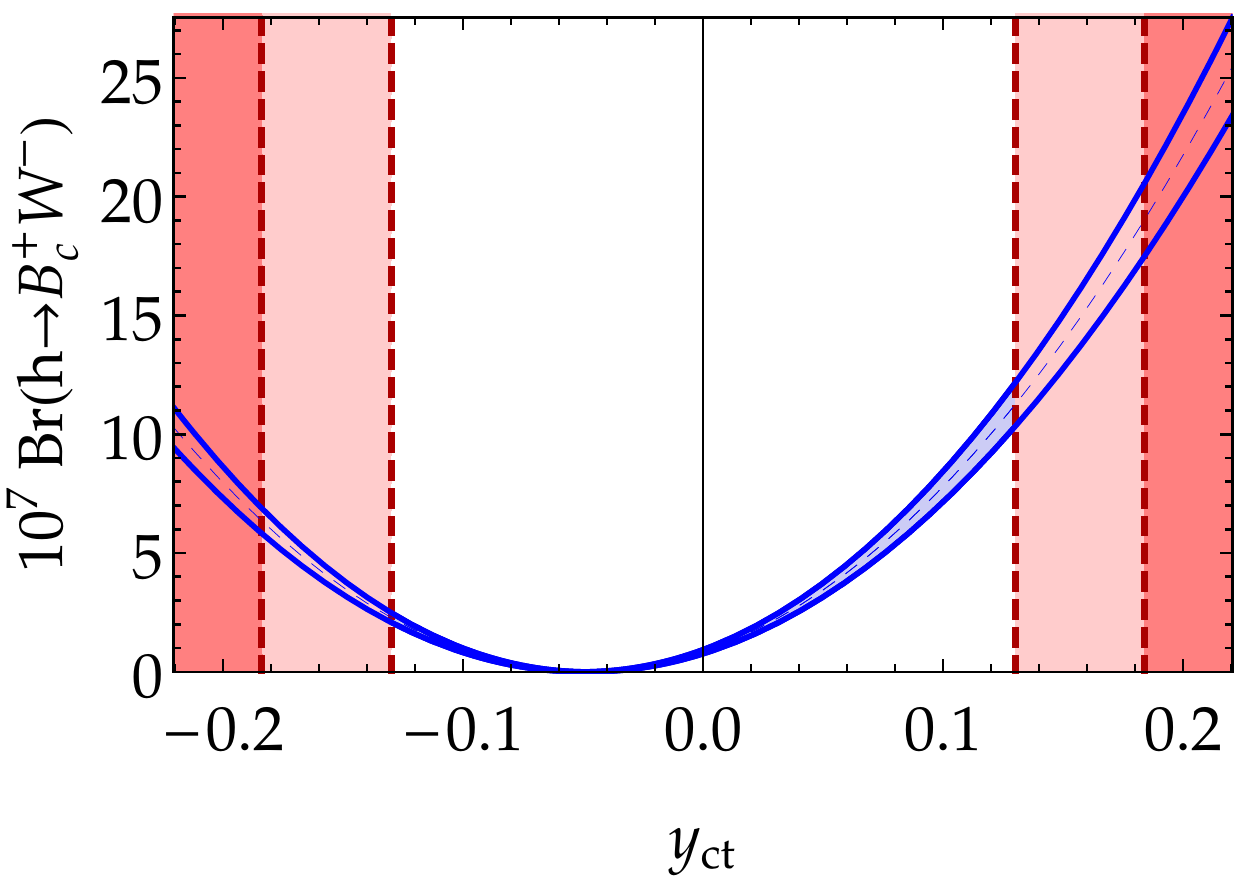}
\parbox{15.5cm}
{\caption{\label{fig:BRB}
Predictions for the branching ratios $\mbox{Br}(h\to B^+ W^-)$ (left) and $\mbox{Br}(h\to B_c^+ W^-)$ (right) as functions of $y_{qt}=\mathrm{Re}(Y_{qt})$. The red bands indicate the exclusion bounds from (\ref{eq:kappaconstr}) when $Y_{tq}=0$ (dark red) and $|Y_{tq}|=|Y_{qt}|$ (bright red).}}
\end{center}
\end{figure}

\begin{figure}
\begin{center}
\includegraphics[width=0.46\textwidth]{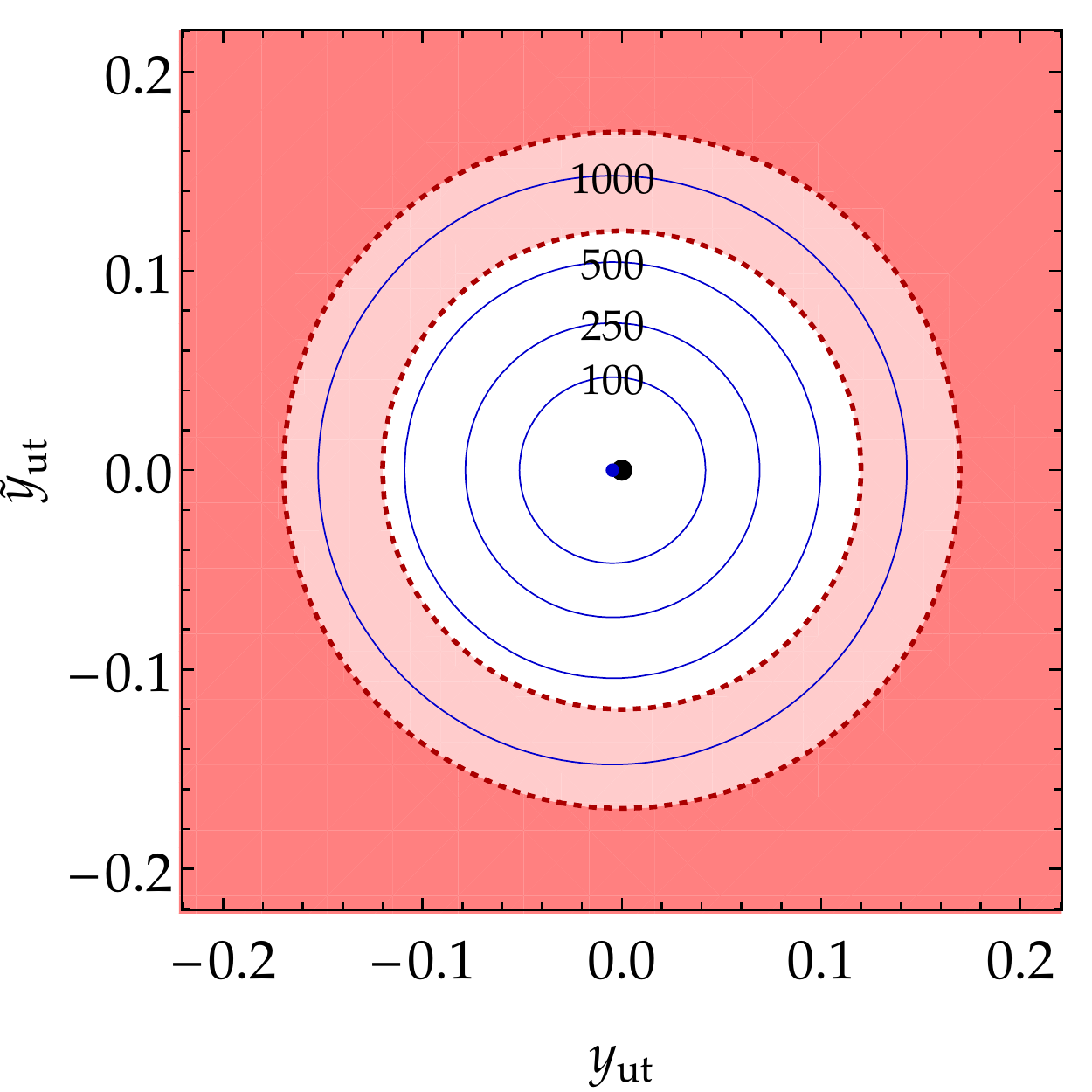}
\quad
\includegraphics[width=0.46\textwidth]{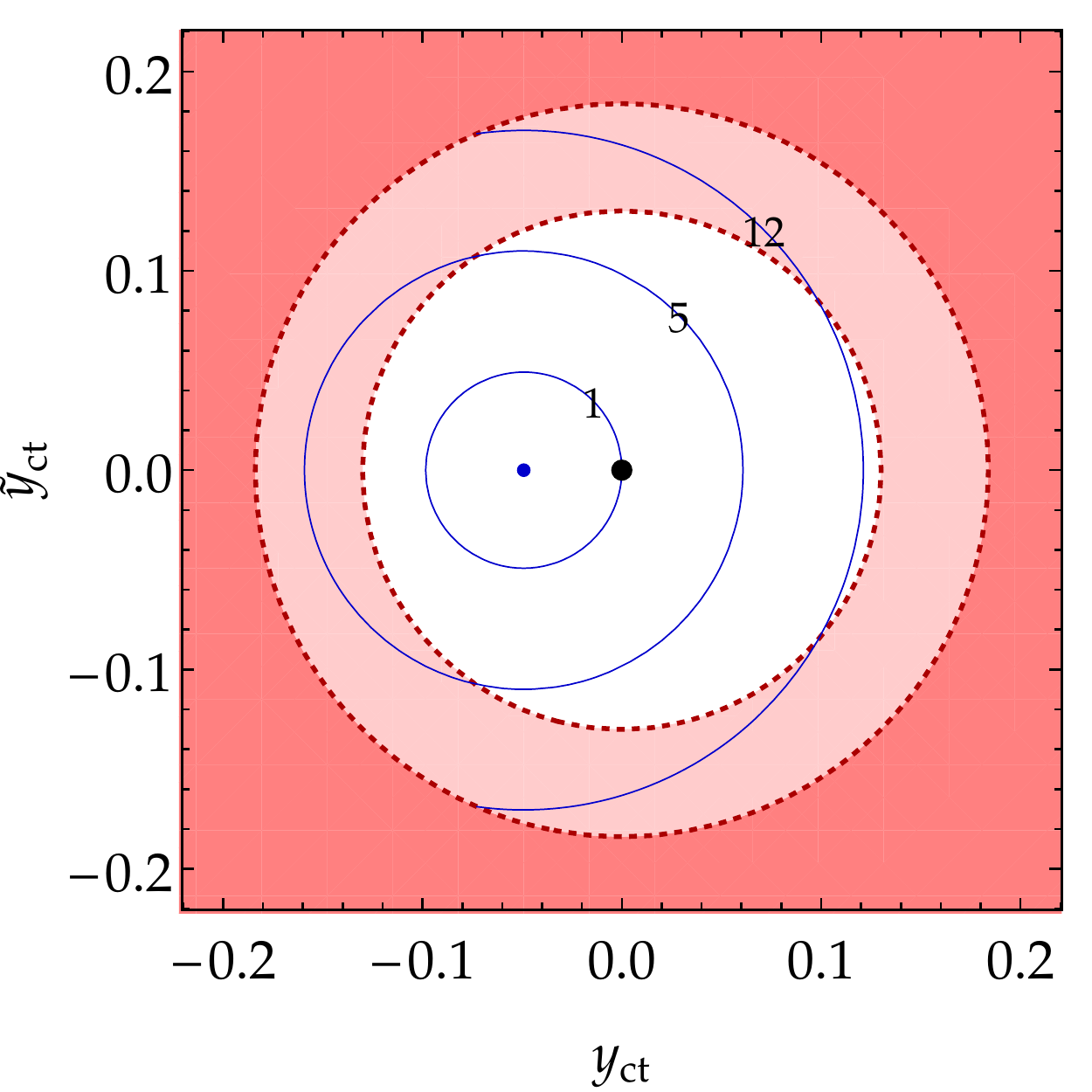}
\parbox{15.5cm}
{\caption{\label{fig:BRBContour}
Enhancement of the branching ratios $\mathrm{Br}(h\to B^{+}W^-)$ (left) and $\mathrm{Br}(h\to B_c^{+} W^-)$ (right) over the SM prediction as functions of the couplings $Y_{qt}=y_{qt}+i\tilde y_{qt}$. The red shaded regions indicate the exclusion bounds from (\ref{eq:kappaconstr}) when $Y_{tq}=0$ (dark red) and $|Y_{tq}|=|Y_{qt}|$ (bright red). The SM value is given by the black dot, while the blue dot denotes the point where the branching ratio vanishes.}}
\end{center}
\end{figure}

The largest effects arise for the decays involving $B$ mesons, since in this case the CKM matrix element $V_{tb}\approx 1$ entering (\ref{eq:FPW-}) and (\ref{eq30}) is unsuppressed. In order to evaluate the direct contributions we adopt the model for the LCDA proposed in \cite{Grossmann:2015lea}, which involves a single width parameter $\sigma_M$. We use $\sigma_B=\sigma_{B^*}=0.087$ and $\sigma_{B_c}=0.305$ at the low hadronic scale $\mu_0=1$\,GeV. We then evolve the LCDA up to the hard scale $\mu_{hW}=(m_h^2-m_W^2)/m_h\approx 73.4$\,GeV. Note that the details of the modelling of the LCDA have a very minor impact on our results, since the integration kernel in the integrals over the LCDAs is a slowly varying function of $x$. In the limit $r_W=1$ the integrals would be determined model-independently by the normalization of the LCDAs. For the case of the $B^*$ vector meson we also need the ratio $f_{B^*}^\perp(\mu)/f_{B^\ast}$. We take the value~1 for this ratio at the low scale $\mu_0$, in accordance with heavy-quark symmetry \cite{Neubert:1993mb}. We then evolve this ratio up to the hard scale $\mu_{hW}$. We can now quote our results for the branching ratios of the decays $h\to B^{(*)+} W^-$ and $h\to B_c^+ W^-$ as functions of the flavor-changing Higgs couplings. We find that the contributions of the direct and the indirect form factor interfere constructively, in contrast to the decays $h\to V\gamma$. We set the off-diagonal Yukawa couplings which do not involve the top quark to zero. We then obtain\footnote{The branching ratio $\mathrm{Br}(h\to B^{*+} W^-)$ has also been calculated in \cite{Kagan:2014ila}. While we agree with their result for the term not involving the off-diagonal Yukawa couplings, we find large deviations in the other terms. Adopting their notation, we find a correction factor $\big[0.98\kappa_W^2+0.02
+7.01\kappa_W\,\mbox{Re}\,\bar\kappa_{ut}-0.45\,\mbox{Re}\,\bar\kappa_{tu}+12.53\,|\bar\kappa_{ut}|^2+5.89\,|\bar\kappa_{tu}|^2\big]$ with respect to the SM, where these authors obtain $\big[\kappa_W^2+26\,\bar\kappa_{ut}^2+22\,\bar\kappa_{tu}^2\big]$.}
\begin{equation}
\begin{aligned}
   \mbox{Br}(h\to B^+ W^-) 
   &= 1.54\cdot 10^{-10} \left( \kappa_W^2 + 427\,\kappa_W\,\mathrm{Re}\,Y_{ut}
    + 45615\,|Y_{ut}|^2 \right) , \\
   \mbox{Br}(h\to B^{*+} W^-) 
   &= 1.41\cdot 10^{-10}\,\big( 0.98\,\kappa_W^2 + 0.02 + 417\,\kappa_W\,\mathrm{Re}\,Y_{ut}
    - 27\,\mathrm{Re}\,Y_{tu} \\
   &\hspace{2.7cm}\mbox{}+ 44296\,|Y_{ut}|^2 + 25833\,|Y_{tu}|^2 \big) \,, \\
   \mbox{Br}(h\to B^+_c W^-) 
   &= 8.21\cdot 10^{-8} \left( \kappa_W^2 + 41\,\kappa_W\,\mathrm{Re}\,Y_{ct} + 413\,|Y_{ct}|^2 \right) .
\end{aligned}
\end{equation}
The CKM-suppression of the indirect contributions combined with the enhancement of the direct contributions described above leads to a strong sensitivity to the flavor-changing Higgs couplings. We demonstrate this dependence in Figure~\ref{fig:BRB}, assuming real couplings $Y_{ut}$ and $Y_{ct}$ and setting $\kappa_W=1$ to its SM value. The bright red band indicates the bound from (\ref{eq:kappaconstr}) one obtains when $|Y_{tq}|=|Y_{qt}|$, while the dark red band corresponds to the assumption that $Y_{tq}=0$. In this most extreme scenario, the $h\to B^+ W^-$ and $h\to B^{*+} W^-$ branching ratios can be enhanced by up to three orders of magnitude with respect to the SM. The enhancement of the $h\to B_c^+ W^-$ branching fraction is less dramatic. Unfortunately, even under the most optimistic assumptions the resulting rates are still predicted to be very small. We demonstrate the full dependence of the ratios $\mbox{Br}(h\to B_{(c)}^+ W^-)$ in the complex plane of $Y_{qt}$ in Figure~\ref{fig:BRBContour}.

\section{Conclusions}
\label{sec:Conc}

We have performed a detailed analysis of the rare exclusive decays $h\to MV$, where $M$ is a pseudoscalar or vector meson and $V=W,Z$ an electroweak gauge boson. The decay amplitudes are governed by two types of amplitude topologies. In the so-called indirect contributions, the Higgs boson couples to the final-state gauge boson $V$ and a second, off-shell gauge boson, which is then converted into the meson $M$. For the case of $h\to MZ$ decays, the off-shell boson can be either a photon or $Z$ boson. While the $hZZ$ coupling exists at tree level in the SM, the $h\gamma Z$ vertex is loop induced and hence suppressed. However, the fact that the photon propagator is almost on-shell counteracts the loop suppression. As a result, the two diagrams are of similar importance, and we find that (with the exception of $h\to\phi Z$) they interfere destructively. The so-called direct contributions to the decay amplitudes involve the Yukawa couplings of the valence quarks in the meson $M$ and are typically subdominant. We have included all three contributions in our theoretical predictions. In the SM, we find $h\to MZ$ branching fractions ranging from $1.5\cdot 10^{-5}$ for $h\to\Upsilon(1S)\,Z$ to $5.6\cdot 10^{-7}$ for $h\to\omega Z$. The $h\to MW$ branching ratios contain the CKM matrix elements corresponding to the final-state mesons $M$. The Cabibbo-allowed modes have branching fractions ranging from $2.5\cdot 10^{-5}$ for $h\to D_s^* W$ to $4.3\cdot 10^{-6}$ for $h\to\pi W$, while CKM-suppressed decay modes have significantly smaller branching ratios.

We have studied the dependence of the branching fractions on physics beyond the SM using an effective Lagrangian, which allows for modifications of the Higgs-boson couplings. The interference pattern of the $h\to MZ$ decay amplitudes mentioned above implies a strong sensitivity to the effective CP-even and CP-odd $h\gamma Z$ couplings. In combination with a future measurement of the $h\to\gamma Z$ decay rate, this can be used to extract these couplings up to a sign ambiguity in the CP-odd coefficient. In the case of the $h\to MW$decay modes, we find an enhanced sensitivity of some of the direct contributions to flavor-changing Higgs couplings involving the top quark. The corresponding decay rates involving $B$ mesons in the final state are strongly CKM suppressed in the SM, but can be significantly enhanced if non-vanishing Yukawa couplings $Y_{qt}$ and $Y_{tq}$ (with $q=u,c$) close to the current experimental upper bounds are assumed. In summary, the rare exclusive Higgs-boson decays explored here exhibit interesting sensitivities to various new-physics effects. This makes them promising targets for precision studies at future experiments like the high-luminosity LHC or a future 100\,TeV proton-proton collider.

\subsubsection*{Acknowledgments}

This work has been supported by the Advanced Grant EFT4LHC of the European Research Council (ERC), the Cluster of Excellence {\em Precision Physics, Fundamental Interactions and Structure of Matter\/} (PRISMA -- EXC 1098), grant 05H12UME of the German Federal Ministry for Education and Research (BMBF), and the DFG Graduate School {\em Symmetry Breaking in Fundamental Interactions\/} (GRK 1581).

\newpage
\begin{appendix}
\renewcommand{\theequation}{A.\arabic{equation}}
\setcounter{equation}{0}

\section{\boldmath Direct contributions to the $h\to MZ$ form factors}
\label{App:SubleadingTwist}

The calculation of the direct contributions to the $h\to MZ$ decay amplitudes is involved, since in many cases the leading terms arise from subleading-twist projections. For pseudoscalar mesons, we use the light-cone projector at leading and subleading twist derived in \cite{Beneke:2000wa,Beneke:2001ev}. Including the leading quark-mass effects, we find 
\begin{equation}\label{eq:A.1}
\begin{aligned}
   F_{\rm direct}^{PZ} 
   &= - \sum_q f^q_P\,\frac{m_q}{2m_h^2} \int_0^1\!dx\,\Bigg\{ 
    \frac{1}{x+r_Z\bar x}\,\bigg[ (a_q \kappa_q-iv_q\tilde \kappa_q)\,m_q\,\phi_P(x) \\
   &\qquad\mbox{}- (a_q \kappa_q+iv_q\tilde \kappa_q)\,\mu_P
    \left( x\,\phi_p(x) + \frac{\phi_\sigma(x)}{3}
    - \Big[ x + \frac{2r_Z}{1-r_Z} \Big]\,\frac{\phi_\sigma'(x)}{6} \right) \bigg] \\
   &\quad\mbox{}+ \frac{1}{\bar x+r_Z x}\,\bigg[ 
    (a_q \kappa_q+iv_q\tilde \kappa_q)\,m_q\,\phi_P(x) \\
   &\qquad\mbox{}- (a_q \kappa_q-iv_q\tilde \kappa_q)\,\mu_P
    \left( \bar x\,\phi_p(x) + \frac{\phi_\sigma(x)}{3}
    + \Big[ \bar x + \frac{2r_Z}{1-r_Z} \Big]\,\frac{\phi_\sigma'(x)}{6} \right) \bigg]
   \\[-1mm]
   &\quad\mbox{}+ \mbox{terms involving 3-particle LCDAs} \Bigg\} \,, 
\end{aligned}
\end{equation}
where for simplicity we omit the scale dependence of the various quantities. For flavor-diagonal final-state mesons, the LCDAs $\phi_P(x)$, $\phi_p(x)$ and $\phi_\sigma(x)$ are symmetric under $x\leftrightarrow (1-x)$, in which case only the terms proportional to $a_q\kappa_q$ survive. At twist-3 order the projector also contains three-particle LCDAs containing a quark, an anti-quark and a gluon. Since the twist-3 LCDAs give strongly suppressed contributions to the decay amplitudes, we will for simplicity neglect the three-particle LCDAs. This is referred to as the Wandzura-Wilczek approximation (WWA) \cite{Wandzura:1977qf}. When this is done, the QCD equations of motion fix the form of the twist-3 LCDAs completely, and one obtains \cite{Braun:1989iv}
\begin{equation}
   \phi_p(x)\big|_{\rm WWA} = 1 \,, \qquad
   \phi_\sigma(x)\big|_{\rm WWA} = 6x(1-x) \,.
\end{equation}
When these expressions are used along with the asymptotic form $\phi_P(x)=6x(1-x)$ of the leading-twist LCDA, one recovers the approximate expressions given in (\ref{FPZdirect}).

The LCDAs of vector mesons at leading and subleading twist have been studied in great detail in \cite{Ali:1993vd,Ball:1996tb,Ball:1998sk}. The corresponding momentum-space projectors were derived in \cite{Beneke:2000wa}. The direct contributions to the form factors $F_{\parallel\,\rm direct}^{VZ}$ for a longitudinally polarized vector meson are obtained from (\ref{eq:A.1}) by making the replacements $f^q_P\to f^q_V$, $v_q\leftrightarrow a_q$, $\phi_P(x)\to\phi_V(x)$, $\mu_P\to m_V f_V^{q\perp}/f_V^q$, and
\begin{equation}
   \phi_p(x)\to \mp\frac12\,h_\parallel^{\prime\,(s)}(x) \,, \qquad
   \frac{\phi_\sigma(x)}{3}\to \pm 2\int_0^x\!dy
    \left[ \phi_V^\perp(y) - h_\parallel^{(t)}(y) \right] , \qquad
   \frac{\phi_\sigma'(x)}{6}\to \mp h_\parallel^{(t)}(x) \,.
\end{equation}
Here the upper (lower) signs refer to the contributions from the first (second) diagram in Figure~\ref{fig:hMZ}, which can be identified by the different denominator structures in (\ref{eq:A.1}). For flavor-diagonal final-state mesons, the LCDAs $\phi_V(x)$ and $h_\parallel^{(t)}(x)$ are symmetric under the exchange $x\leftrightarrow(1-x)$, while $h_\parallel^{\prime\,(s)}(x)$ and $\int_0^x\!dy\big[\phi_V^\perp(y)-h_\parallel^{(t)}(y)\big]$ is anti-symmetric. In this case only the terms proportional to $v_q \kappa_q$ survive. In the approximation where three-particle LCDAs are neglected, the QCD equations of motion imply the relations \cite{Ball:1998sk,Beneke:2000wa}
\begin{equation}
\begin{aligned}
   h_\parallel^{(t)}(x,\mu) \big|_{\rm WWA} &= (2x-1)\,\Phi_v(x,\mu) \,, \qquad
    h_\parallel^{\prime\,(s)}(x,\mu) \big|_{\rm WWA} = -2\Phi_v(x,\mu) \,, \\
   &\hspace{-12mm} \int_0^x\!dy \left[ \phi_V^\perp(y,\mu) - h_\parallel^{(t)}(y,\mu) 
    \right]_{\rm WWA} = x(1-x)\,\Phi_v(x,\mu) \,,
\end{aligned}
\end{equation}
where
\begin{equation}
   \Phi_v(x,\mu) = \int_0^x\!dy\,\frac{\phi_V^\perp(y,\mu)}{1-y} 
    - \int_x^1\!dy\,\frac{\phi_V^\perp(y,\mu)}{y} \,.
\end{equation}
In this approximation, the twist-3 two-particle amplitudes can be expressed in terms of the leading-twist LCDA $\phi_V^\perp$. When the asymptotic form $\phi_V^\perp(x)=6x(1-x)$ is used, we find
\begin{equation}
   F_{\parallel\,{\rm direct}}^{VZ} 
   = -3 \sum_q v_q\,\kappa_q\,\frac{m_q}{m_h^2} \left[
    4r_Z f_V^{q\perp} m_V\,\frac{2(1-r_Z)+(1+r_Z)\ln r_Z}{(1-r_Z)^4}
    + f_V^q\,m_q\,\frac{1-r_Z^2+2r_Z\ln r_Z}{(1-r_Z)^3} \right] \!.
\end{equation}

For transversely polarized vector mesons, the direct contributions to the form factors arise from leading-twist projections. We find
\begin{equation}\label{F1dirVZ}
\begin{aligned}
   F_{\perp\,{\rm direct}}^{VZ} 
   &= \frac{1-r_Z}{4}\,\sum_q \frac{m_q f_V^{q\perp}}{m_V}
    \int_0^1\!dx \left( \frac{v_q \kappa_q+ia_q\tilde \kappa_q}{x+r_Z\bar x}
    + \frac{v_q \kappa_q-ia_q\tilde \kappa_q}{\bar x+r_Z x} \right) \phi_V^\perp(x) \,, \\
   \tilde F_{\perp\,{\rm direct}}^{VZ} 
   &=\,\frac{1-r_Z}{4}\,\sum_q \frac{m_q f_V^{q\perp}}{m_V}
    \int_0^1\!dx \left( \frac{v_q\tilde \kappa_q-ia_q \kappa_q}{x+r_Z\bar x}
    + \frac{v_q\tilde \kappa_q+ia_q \kappa_q}{\bar x+r_Z x} \right) \phi_V^\perp(x) \,.
\end{aligned}    
\end{equation}
For flavor-diagonal final-state mesons, the LCDA $\phi_V^\perp(x)$ is symmetric under $x\leftrightarrow(1-x)$, and hence only the terms proportional to $v_q \kappa_q$ survive. When the asymptotic form $\phi_V^\perp(x)=6x(1-x)$ is used, one recovers the approximate expressions given in (\ref{Ftransdirect}).

We finally quote the generalization of relation (\ref{eq22}), valid for flavor-changing decays involving vector mesons containing different quark flavors $q$ and $q'$. At leading-twist order, we obtain
\begin{equation}
\begin{aligned}
   &\Gamma(h\to V_{qq'} Z) 
   = \frac{m_h  (f_V^\perp)^2}{64\pi v^2}\,r_Z (1-r_Z)^3 \\
   &\times \left\{ \left| \int_0^1\!dx \left(
    \frac{v_q (Y_{qq'}\!+\!Y^*_{q'q}) + a_q (Y_{qq'}\!-\!Y_{q'q}^*)}{x+r_Z\bar x}
    + \frac{v_{q'} (Y_{qq'}\!+\!Y_{q'q}^* ) - a_{q'} (Y_{qq'}\!-\!Y_{q'q}^*)}{\bar x+r_Z x} \right) 
    \phi_V^\perp(x) \right|^2 \right.\\
   &\hspace{3mm}\mbox{}+ \left. \left| \int_0^1\!dx \left(
    \frac{v_q (Y_{qq'}\!-\!Y_{q'q}^*) + a_q (Y_{qq'}\!+\!Y_{q'q}^*)}{x+r_Z\bar x}
    + \frac{v_{q'}(Y_{qq'}\!-\!Y_{q'q}^*) - a_{q'}(Y_{qq'}\!+\!Y_{q'q}^*)}{\bar x+r_Z x} \right) 
    \phi_V^\perp(x) \right|^2\right\} .
\end{aligned}
\end{equation}
This expression reduces to (\ref{eq22}) when the asymptotic form $\phi_V^\perp(x)=6x(1-x)$ is employed.

\end{appendix}

\end{document}